%% file: bare_conf.tex
\documentclass[conference]{IEEEtran}

\usepackage[OT1]{fontenc}
\usepackage[utf8]{inputenc}

\usepackage[cmex10]{amsmath}
\usepackage{bm}
\usepackage{amsfonts}
\usepackage{amssymb}
\usepackage{pgfplots}
\pgfplotsset{compat=newest}
\usepgfplotslibrary{groupplots}
\usepackage{tikz}
\usetikzlibrary{matrix, positioning, patterns, shapes, arrows, spy}
\usepackage{xcolor}
\definecolor{Paired-2}{RGB}{166,206,227}
\definecolor{Paired-1}{RGB}{31,120,180}
\definecolor{Paired-4}{RGB}{178,223,138}
\definecolor{Paired-3}{RGB}{51,160,44}
\definecolor{Paired-6}{RGB}{251,154,153}
\definecolor{Paired-5}{RGB}{227,26,28}
\definecolor{Paired-8}{RGB}{253,191,111}
\definecolor{Paired-7}{RGB}{255,127,0}
\definecolor{Paired-10}{RGB}{202,178,214}
\definecolor{Paired-9}{RGB}{106,61,154}
\definecolor{Paired-12}{RGB}{255,255,153}
\definecolor{Paired-11}{RGB}{177,89,40}

\DeclareMathOperator{\sign}{sign}

\usepackage{subcaption}
\usepackage{cleveref}

\usepackage{booktabs}
\usepackage[tworuled, linesnumbered, vlined]{algorithm2e}
\usepackage{cuted}
\usepackage{soul} 

\begin{document}
\bstctlcite{IEEEexample:BSTcontrol}
\title{Asymmetric Construction of Low-Latency and Length-Flexible Polar Codes}

\author{\IEEEauthorblockN{Adam Cavatassi, Thibaud Tonnellier and Warren J. Gross}
\IEEEauthorblockA{Department of Electrical and Computer Engineering\\
McGill University, Montr\'eal, Qu\'ebec, Canada\\
Email: adam.cavatassi@mail.mcgill.ca, thibaud.tonnellier@mail.mcgill.ca, warren.gross@mcgill.ca}}

\maketitle

\begin{abstract}
Polar codes are a class of capacity-achieving error correcting codes that have been selected 
for use in enhanced mobile broadband in the 3GPP 5\textsuperscript{th} generation (5G) wireless standard. Most polar code research examines 
the original Ar{\i}kan polar coding scheme, which is limited in block length to powers of two. This constraint presents
a considerable obstacle since practical applications call for all code lengths to be readily available.
Puncturing and shortening techniques allow for flexible polar codes, while multi-kernel polar codes
produce native code lengths that are powers of two
and/or three. In this work, we propose a new low complexity coding scheme called asymmetric
polar coding that allows for any arbitrary block length. We present details on the generator matrix, 
frozen set design, and decoding schedule. Our scheme offers flexible polar code lengths with
decoding complexity lower than equivalent state-of-the-art length-compatible approaches under successive 
cancellation decoding. Further, asymmetric decoding complexity is directly dependent on the codeword length rather than the nearest valid polar code length. 
We compare our scheme with other length matching techniques, and simulations are presented. Results show that asymmetric polar codes
present similar error correction performance to the competing schemes, while dividing the number of SC decoding
operations by up to a factor of 2 using the same codeword length.
\end{abstract}

\IEEEpeerreviewmaketitle

\section{Introduction}
\label{sec:intro}
Polar codes are capacity achieving error-correcting codes \cite{Arikan2009} that have been selected for
control channel use in the 3GPP 5th generation (5G) New Radio standard \cite{3GPP}.
Conventional (or Ar{\i}kan) polar codes can natively only attain length $N$ such that $N=2^n$ 
with $n\in \mathbb{N}^+$. The generator matrix of an Ar{\i}kan polar code is obtained 
by applying the $n$-th Kronecker product, denoted as $\otimes$, on the Ar{\i}kan kernel 
$\bm{T_2} = \big[\begin{smallmatrix} 1 & 0 \\ 1 & 1 \end{smallmatrix}\big]$, 
resulting in an $N \times N$ matrix. However, length-flexible codes are 
mandatory for practical communication systems. LDPC codes were favoured for the data channel in 3GPP New Radio
due to their length flexibility and their decoding complexity, which is directly linked to their codeword size \cite{Richardson2018}.
Thus far, multiple promising schemes have been proposed
to assuage the length restriction of polar codes: puncturing \cite{Niu2013} and shortening  \cite{Wang2014} (PS) and multi-kernel (MK) polar codes \cite{Gabry2016}. 
Chained polar subcodes \cite{7938042} were shown to be an effective method
for achieving length compatible polar codes, although their complexity precludes their practicality
and thus are out of the scope of this paper.

Several PS schemes have been introduced whereby
the encoding of a block length of size $N_P$ is realized by considering 
a length $2^{N_M}$ mother polar code where $N_M = \lceil \log_2N_P \rceil$. Then, $2^{N_M} - N_P$ 
bits are either shortened or punctured and are not transmitted over the channel. Decoding is then enacted on the mother code. 
PS patterns have an impact on bit reliabilities, and so code construction must be co-optimized 
with the PS sets. Such complications can be neglected with the low complexity methods 
proposed in \cite{Bioglio2017}. Furthermore, the decoding computational complexity of 
PS schemes is dependent on that of the mother code .

MK codes \cite{Gabry2016}, \cite{Benammar2017} are a technique that combines
differently sized kernels. 
Although larger sizes are possible, kernels sizes 2 and 3 are most common, allowing any block 
length $N_{MK} = 2^n3^m$ to be obtained. MK codes offer 
improved native length flexibility over Ar{\i}kan polar codes, though they 
introduce further complexity to decoding and code design \cite{Coppolino2018}.

In this paper, we propose a new coding scheme in which polar codes of unequal lengths 
are linked together using polar transformations. We refer to the new codes as asymmetric 
polar codes (APC). By linking multiple polar codes together, any arbitrary block length can be 
achieved. Our method offers a straightforward approach to polar coding that is length-flexible and 
exhibits length-dependent decoding complexity. 
APCs are similar to Ar{\i}kan polar codes in the sense that they both 
have a recursive structure and contain smaller polar codes in their generator matrix. 
Many previously designed polar decoders and construction methods 
for Ar{\i}kan codes are applicable to APCs. 
Notably, APCs feature reduced time and space complexity with comparable frame error rate (FER) 
performance to equivalent PS codes. APCs also have inherently reduced decoding latency against PS codes due to the sequential nature of successive cancellation.

The remainder of this paper is organized as follows: Section II reviews polar code preliminaries, and existing length 
compatible polar code principles. In Section III, asymmetric polar codes are 
introduced, including details on encoding, decoding, and code construction. In Section IV 
experimental results are provided both in terms of decoding performance and 
complexity evaluation.

\section{Polar Codes}
\label{sec:pc}

\subsection{Generator Matrix Construction}
\label{sec:pc:gen_matrix}
A polar code, denoted by $\mathcal{PC}(N,K)$, is a linear block code of length $N$ and rate $R = \frac{K}{N}$. Channel polarization causes
individual bit indices to act as synthetic subchannels with either a higher or lower reliability than the original uncoded channel \cite{Arikan2009}.
Among the $N$ sub-channels, the first $K$ indices from $\mathcal{R}$, an index vector sorted by reliability, constitute the information set $\mathcal{I}$ to transmit data. The $N-K$ remaining positions 
comprise the frozen set $\mathcal{F}$ and are typically set to 0. 

The uncoded input vector can be written as $\bm{u} = (u_0, u_1, \hdots, u_{N-1})$, and the polar codeword
is obtained with $\bm{x} = \bm{u}\bm{G_N}$, where the generator matrix $\bm{G_N} = \bm{T_2}^{\otimes n}$. A key property of polar codes is that the encoder is recursive. As such, the generator 
matrix of a code of length $N=2^n$ contains in it all smaller polar codes of length $N=2^{n-1}, 2^{n-2},$ etc. Codeword $\bm{x}$ is transmitted over channel $W$ and received 
as vector of log likelihood ratios (LLR) $\bm{y}$. References to \textit{Ar{\i}kan} polar codes in this paper will indicate this construction.

\subsection{Decoding}
\label{sec:pc:decode}
The fundamental decoder for polar codes is known as successive cancellation (SC), which was proposed in \cite{Arikan2009}. It can be viewed as
binary tree traversal with left-node-first priority. The noisy codeword $\bm{y}$ is the decoder input at the top of the tree, corresponding to tree stage $d = n$. 
Each node $v$ contains $N_v=2^{d}$ LLRs $\bm{\alpha}$ and bit partial sums $\bm{\beta}$. At each node, the left and right children LLRs, 
$\bm{\alpha}^l$ and $\bm{\alpha}^r$, are computed using $f$ and $g$, respectively:
\vspace{-.5\baselineskip}

\small
\begin{equation}
\label{eq:f_g}
	\begin{aligned}
		\alpha_i^l &= f(\alpha_i,\alpha_{i+\frac{N_v}{2}}) = \alpha_i \boxplus \alpha_{i+\frac{N_v}{2}}, \\
		\alpha_i^r &= g(\alpha_i,\alpha_{i+\frac{N_v}{2}},\beta_{i}^l) = (-1)^{\beta_{i}^l} \cdot \alpha_i + \alpha_{i+\frac{N_v}{2}},
	\end{aligned}
	\quad \forall i \in [0,\frac{N_v}{2}-1]
\end{equation}
\normalsize
where $a \boxplus b \approx \sign{(a)}\sign{(b)}\min{(\lvert a \rvert,\lvert b \rvert)}$.
The $HD$ function is used for computing the bit decision at each leaf node as: 
\begin{equation}
\label{eq:hd}
		\hat{u}_i=HD(\alpha_i) = \begin{cases} 0 & \text{if } \alpha_i > 0 \text{ or } i \in \mathcal{F} \\ 
								1 & \text{otherwise}.
				 	\end{cases}
\end{equation}
After returning from a right branch, the partial sums in the parent node are updated with 
$h(\beta_{i},\beta_{i+\frac{N_v}{2}}) = (\beta_{i}^l \oplus \beta_{i}^r, \beta_{i}^r)$ before exploring another branch. Ostensibly, an SC decoder can be reduced to a schedule of 
$f$ and $g$ functions, where the total number of $f$ and $g$ operations is given by $N \log_2{N}$.

To overcome mediocre error rate performance of SC, SC List (SCL) has been proposed in \cite{Tal2015}. In the SCL case, each leaf node considers
both possible bit values (0 and 1). $L$ different decoding paths are maintained using a candidate competition based on a path metric, which
dictates the pruned paths. SCL becomes more powerful when a CRC (SCL-CRC) is concatenated before polar encoding to aid in path candidate selection. 
SC and SCL are further enhanced by latency reduction using Fast-SSC \cite{Sarkis2014} \cite{Sarkis2016}, 
which exploits frozen bit locations to identify specialized nodes. 

\subsection{Punctured and Shortened Polar Codes}
\label{sec:pc:ps}
The standard methods for attaining length compatible polar codes are known as puncturing \cite{Niu2013}
and shortening \cite{Wang2014}. To build a polar code of arbitrary length $N_S$, both methods
involve freezing additional bits of a larger polar code of length $N_M = 2^{\lceil \log_2{N_S} \rceil}$ to 
rectify the length difference. This \textit{mother} polar code is used for encoding and decoding. 
In the case of shortening, $S = N_M - N_S$ bits in $\bm{u}$ are set to $0$,
such that the corresponding indices in $\bm{x}$ are known to be $0$. These indices are said to be \textit{overcapable} 
and are added to the shortening set $\bm{S}$. To encode, $\bm{S}$ must be included in $\mathcal{F}$.
Indices in $\bm{S}$ are not transmitted over the channel, and
$\bm{S}$ must be known at the decoder. Since shortened indices are known to be 0, their LLRs are set to infinity, 
or a sufficiently large value in practice, prior to decoding.

Similarly, puncturing $P =  N_M - N_P$ bits allows construction of a code of length $N_P < N_M$ from a mother
code. $\bm{P}$ contains $P$ \textit{incapable} indices that are added to $\mathcal{F}$
and not transmitted. Incapable indices are those that are completely unreliable at their leaf node when the corresponding 
channel value is unreliable during decoding, \emph{ie.} $\alpha_i=0$ when $y_i = 0$ \cite{Bioglio2017}.
LLRs of these indices are treated as erasures and set to $0$ before the decoder is enacted. 

The Wang-Liu (WL) \cite{Wang2014} method for shortening and quasi-uniform
puncturing (QUP) \cite{Niu2013} produce frozen sets that weigh the modified reliabilities of 
punctured and shortened bits. As such,  
these methods require reliability computation after the sets $\bm{P}$ or $\bm{S}$ are decided. 
The frozen set must then be redesigned for every length. Bit reversal schemes were proposed to eliminate 
the optimization overhead of WL/QUP and produce good results \cite{Bioglio2017}. 
Shortening typically exhibits superior performance to puncturing in high code rates, while the opposite is true for low code rates.
In this paper, we will only consider the WL and QUP methods in our evaluation, as these methods are generally superior to the low-complexity 
methods in terms of error correction performance \cite{Bioglio2017}. 

\subsection{Multi-Kernel Polar Codes}
\label{sec:pc:mk}
MK polar codes were introduced in \cite{Gabry2016} as an alternative polarizing construction to attain
polar codes of lengths other than powers of $2$. The ternary kernel $\bm{T_3} = \bigg[\begin{smallmatrix} 1 & 1 & 1 \\ 1 & 0 & 1 \\ 0 & 1 & 1 \end{smallmatrix}\bigg]$,
was proposed and shown to be optimal in its polarizing effects \cite{Benammar2017}. It can be used as a Kronecker product component in
conjunction with the Ar{\i}kan kernel, to produce polar codes of length $N_{MK} = 2^n3^m \text{ for } n,m\in \mathbb{N}$.
The native length flexibility of polar codes is thus improved, however encoding and decoding becomes more convoluted.
SC decoding is modified in ternary stages by computing three branches $\bm{\alpha}^l$, $\bm{\alpha}^c$, and $\bm{\alpha}^r$ for each node:
\begin{equation}
\vspace{-.4\baselineskip}
\label{eq:mk_dec}
	\begin{aligned}
		\alpha_i^l &= \alpha_i \! \boxplus  \! \alpha_{i+\frac{N_v}{3}} \! \boxplus  \! \alpha_{i+\frac{2N_v}{3}}, \\
		\alpha_i^c &= (-1)^{\beta_{i}^l} \!\cdot\! \alpha_i \!+\! \alpha_{i+\frac{N_v}{3}} \!\boxplus\! \alpha_{i+\frac{2N_v}{3}}, \\
		\alpha_i^r &= (-1)^{\beta_{i}^l} \!\cdot\! \alpha_{i+\frac{N_v}{3}} \!+\! (-1)^{\beta_{i}^l \otimes \beta_{i}^c} \!\cdot\! \alpha_{i+\frac{2N_v}{3}}. 
	\end{aligned}
	\vspace{-.\baselineskip}
\end{equation}
$\bm{\beta}$ in ternary stages are updated with $h_3(\beta_{i},\beta_{i+\frac{N_v}{3}},\beta_{i+\frac{2N_v}{3}}) = (\beta_{i}^l \oplus \beta_{i}^c, \beta_{i}^l \oplus \beta_{i}^r, \beta_{i}^l \oplus \beta_{i}^c \oplus \beta_{i}^r)$.
MK codes introduce complexity by compelling optimization and storage of the kernel order. For example, a code of length $N_{MK} = 6$ can be built
with either Kronecker products $\bm{T_2} \otimes \bm{T_3}$ or $\bm{T_3} \otimes \bm{T_2}$, which will result in two unique generator matrices.
The number of SC decoding operations for MK codes is given by $N_{MK}(n+m)$. 
All instances of MK codes for the remainder of this paper refer to polar code constructions utilizing $\bm{T_2}$ and $\bm{T_3}$ only.

\section{Low Latency Length Compatible Polar Codes}
\label{sec:apc}
In this section, we propose a new flexible polar coding scheme called asymmetric polar codes. 
Commonly researched topics in polar codes such as decoders, construction methods, or hardware 
implementations can apply to APCs with only some modifications in scheduling.  

\begin{figure}[t]
\makebox[\linewidth][c]{
	\centering
	\hspace{7mm}
	\begin{subfigure}[t]{.21\textwidth}
	\centering
		\includegraphics[scale = 0.29]{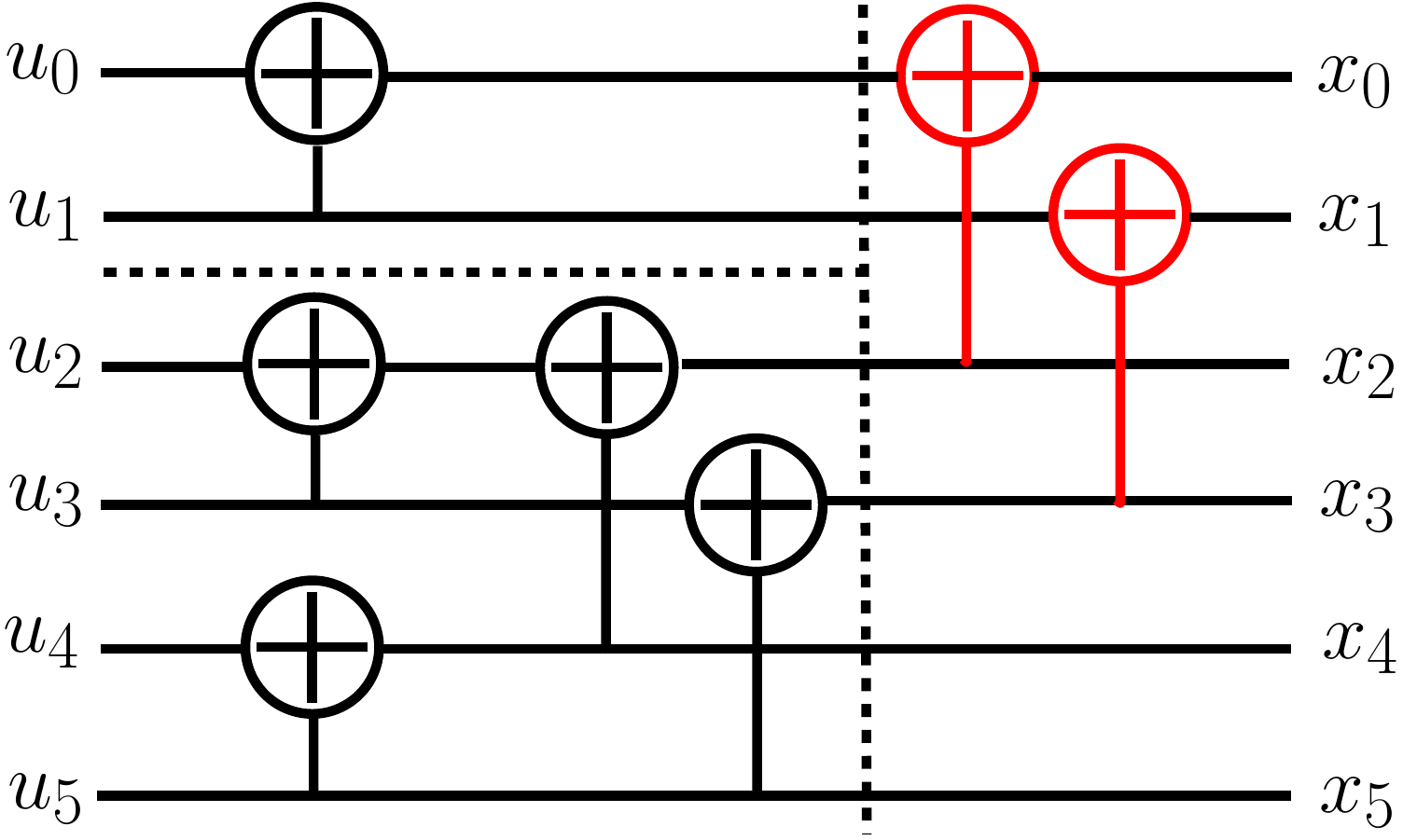}
		\caption{Encoder}
		\label{fig:n_6:enc:asc}
	\end{subfigure}
	\begin{subfigure}[t]{.3\textwidth}
	\centering
		\includegraphics[scale = 0.27]{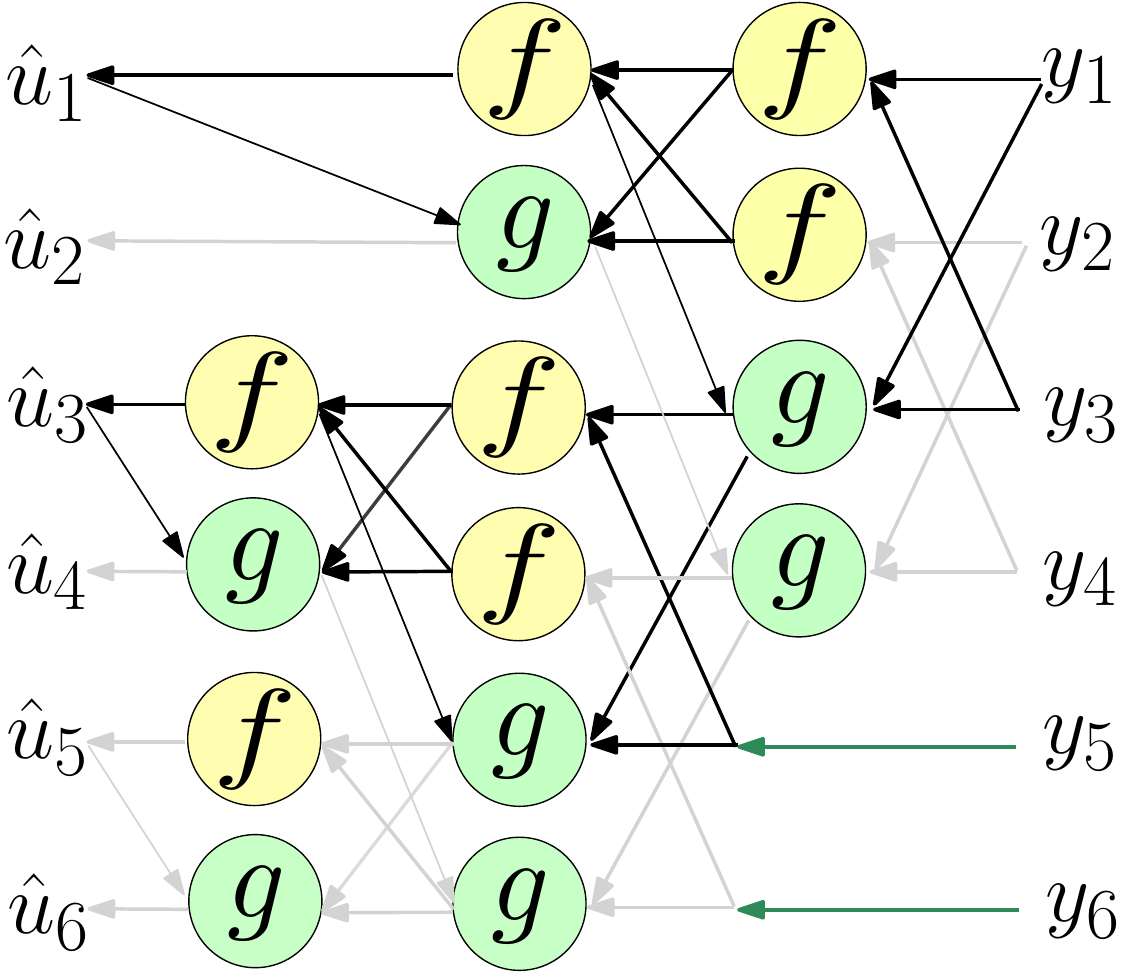}
		\caption{Decoder}
		\label{fig:n_6:dec:asc}
	\end{subfigure}
	}
	\caption{An \textit{ascending} APC of length $N_A=6$ where $\bm{N} = \{4,2\}$ and $p=2$.}
	\vspace{-.5\baselineskip}
\end{figure}

\begin{figure}[t]
\makebox[\linewidth][c]{
	\centering
	\hspace{7mm}
	\begin{subfigure}[t]{.21\textwidth}
	\centering
		\includegraphics[scale = 0.29]{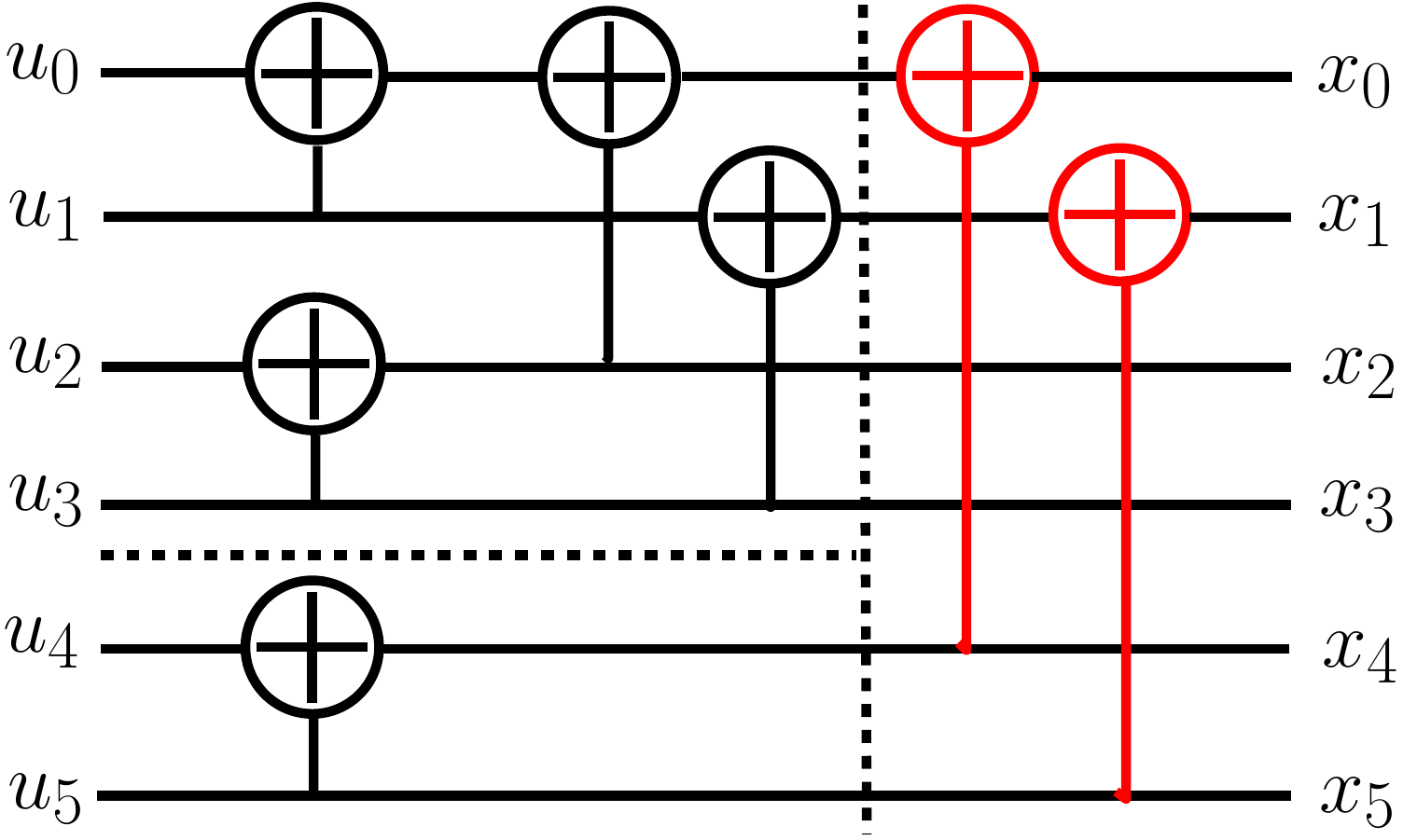}
		\caption{Encoder}
		\label{fig:n_6:enc:dec}
	\end{subfigure}
	\begin{subfigure}[t]{.3\textwidth}
	\centering
		\includegraphics[scale = 0.27]{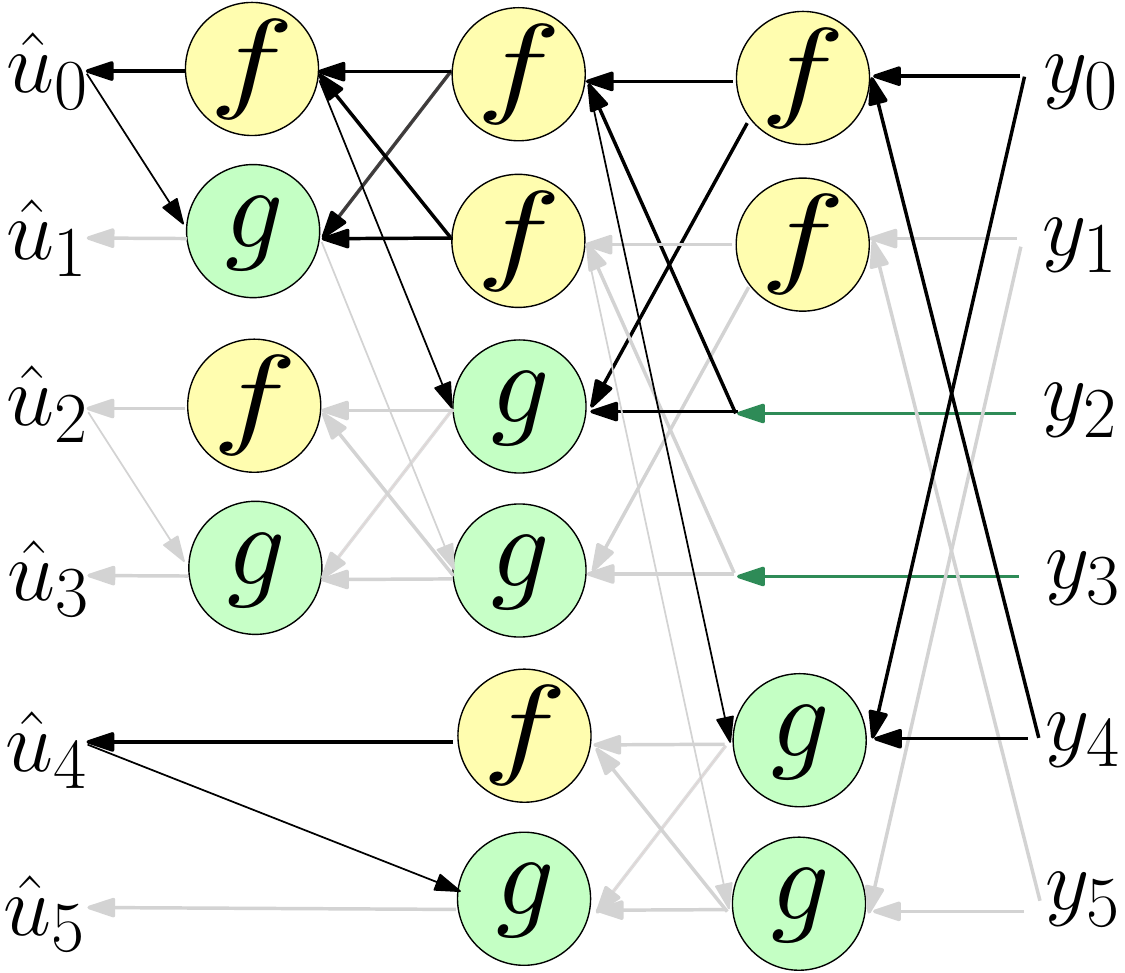}
		\caption{Decoder}
		\label{fig:n_6:dec:dec}
	\end{subfigure}
	}
	\caption{A \textit{descending} APC of length $N_A=6$ where $\bm{N} = \{2,4\}$ and $p=2$.}
	\vspace{-1\baselineskip}
\end{figure}

\subsection{Generator Matrix Construction}
\label{sec:apc:gen_matrix}
An APC of length $N_A$ is constructed from a minimum $p$ partial polar codes, which are determined by the 
decomposition of $N_A$ into a sum of powers of 2, which encompasses the vector $\bm{N} = \{N_0, N_1, \hdots, N_{p-1}\}$.
For example, a code length of $N_A=14$ is represented in binary
as $1110_2$, and therefore $\bm{N} = \{8, 4, 2\}$ and $p=3$. 
$\bm{N}$ can be ordered so that APCs be constructed in either an \textit{ascending} ($asc=1$) or \textit{descending} ($asc=0$) permutation. 
The ascending order indicates that the size of the partial codes increases with bit indices, as in Fig. \ref{fig:n_6:enc:asc}.
Alternatively, the descending order follows that the partial code sizes decrease with bit indices, which requires that $\bm{N}$
be reversed, as in Fig. \ref{fig:n_6:enc:dec}. 
The linking process is 
executed recursively whereby each partial code $G_{N_l}$ is linked with the next partial code $G_{N_{l+1}}$ according to the sequence $\bm{N}$. 
Generator matrix assembly for APCs requires $p-1$ iterations of this process. The polarizing stage 
used for each iteration $l$ is the same XOR operation that is used for the last stage of an Ar{\i}kan code of length $2N_{l+1}$. 
This stage will include $J_l=\min(\sum_{k=0}^{l+1}N_k-N_{l+1}, N_{l+1})$ sum junctions, detailed in Fig. \ref{fig:const_n7:junction}, which join indices $j$ and $j+N_{l+1}$ for $j \in [0,J_l-1]$.
The generator matrix $G_{N_A}$ is equal to the final iteration of the linking 
matrix, such that $G_{N_A} = L_{p-2}$, where $L_l$ is defined in eq. \ref{eq:apc:gen_matrix}.
\begin{figure}
\begin{equation}
\label{eq:apc:gen_matrix}
		L_l = 
			\begin{cases}
			\hfil G_{N_l} & l = 0 \vspace{2mm} \\
			\left[
				\begin{array}{c|c}
					G_{N_l} & \bm{0} \\
			\hline	G_{N_l} \otimes \bm{\hat{1}_{\frac{\sum_{j<l}{N_j}}{N_l}}^{\intercal} } & L_{l-1}
				\end{array} \right] & l > 0 , ~asc = 1 \\[14pt]

			\left[
				\begin{array}{c|c}
					G_{N_l} & \bm{0} \\
			\hline	G_{N_l}[\sum_{j<l}{N_j} ;] & L_{l-1}
				\end{array}  \right] & l > 0 , ~asc = 0,

			\end{cases}
\end{equation}
\vspace{-1.5\baselineskip}
\end{figure}
where $\bm{\hat{1}}_k$ is the all-one vector of length $k$, and $\bm{G}[a;]$ is generator matrix $\bm{G}$ with only its first $a$ rows. 
It should be noted that $\bm{N}$ can contain a code length of $1$, 
which equates to a single uncoded bit, \emph{ie.} $G_1 \!=\! \begin{bmatrix} 1 \end{bmatrix}$. Observe that the linking matrix closely 
resembles the Ar{\i}kan kernel where $l>0$. Further, this representation makes it possible to obtain Ar{\i}kan polar codes 
when $\bm{N}$ contains only two equal lengths. We will only consider the two permutations of $\bm{N}$ using the minimum required 
partial polar codes in our analysis.

\subsection{Example: $N_A = 6$}
\label{sec:apc:ex}
Observe in Fig. \ref{fig:n_6:enc:asc} that an encoder with $N_A = 6$ has constituent polar codes of length 2 and 4. 
Combining these two partial codes is done with an additional polarizing stage. The additional stage is the same as the last stage
of an Ar{\i}kan polar code of length twice that of the upper code in the Tanner graph. In particular, the upper code in this example has length 2 
and thus requires the last stage of a polar code of length 4. The SC decoding schedule must be modified to match the new Tanner graph, as in Fig. \ref{fig:n_6:dec:asc}.

An APC generator matrix $\bm{G_{N_A}}$ must contain those of the constituent codes so as to be consistent 
with the original polar code definition. Thus, $\bm{G_{N_A}}$ is a block
matrix comprised of the partial code matrices and the additional polarizing transforms
discussed above. In this case, the final ascending matrix $\bm{G_6}^{\text{ASC}}$ in eq. (\ref{eq:ex_build_g6}) has block components $\bm{G_2}$ (upper left) and $\bm{G_4}$ (lower right) that follow the $\bm{G_N}$ definition
in Section \ref{sec:pc:gen_matrix}. Similarly, the descending matrix $\bm{G_6}^{\text{DES}}$ is also presented in eq. (\ref{eq:ex_build_g6}).
\begin{equation}
  \label{eq:ex_build_g6}
  \begin{array}{cc}
          \bm{G_{6}}^{\text{ASC}} = 
        \left[
        \setlength\arraycolsep{2pt} 
        \def\arraystretch{0.85}
          \begin{array}{cc|cccc}
          \color{red}1 & \color{red}0 & 0 & 0 & 0 & 0 \\
          \color{red}1 & \color{red}1 & 0 & 0 & 0 & 0 \\
      \hline  1 & 0 & \color{Paired-3}1 & \color{Paired-3}0 & \color{Paired-3}0 & \color{Paired-3}0 \\
          1 & 1 & \color{Paired-3}1 & \color{Paired-3}1 & \color{Paired-3}0 & \color{Paired-3}0 \\
          1 & 0 & \color{Paired-3}1 & \color{Paired-3}0 & \color{Paired-3}1 & \color{Paired-3}0 \\
          1 & 1 & \color{Paired-3}1 & \color{Paired-3}1 & \color{Paired-3}1 & \color{Paired-3}1 \\
          \end{array}
        \right]
        &
      \bm{G_{6}}^{\text{DES}} = 
        \left[
        \setlength\arraycolsep{2pt}
        \def\arraystretch{0.85}
          \begin{array}{cccc|cc}
          \color{Paired-3}1 & \color{Paired-3}0 & \color{Paired-3}0 & \color{Paired-3}0 & 0 & 0 \\
          \color{Paired-3}1 & \color{Paired-3}1 & \color{Paired-3}0 & \color{Paired-3}0 & 0 & 0 \\
          \color{Paired-3}1 & \color{Paired-3}0 & \color{Paired-3}1 & \color{Paired-3}0 & 0 & 0 \\
          \color{Paired-3}1 & \color{Paired-3}1 & \color{Paired-3}1 & \color{Paired-3}1 & 0 & 0 \\
 \hline   1 & 0 & 0 & 0 & \color{red}1 & \color{red}0 \\
          1 & 1 & 0 & 0 & \color{red}1 & \color{red}1 \\
          \end{array}
        \right]
    \end{array}
\end{equation}

\subsection{Frozen Set Construction}
\label{sec:apc:frozen}
\begin{figure}[t]{}
\vspace*{.5\baselineskip}
\makebox[\linewidth][c]{{}
	\centering
	\hspace{-4mm}
	\begin{subfigure}[]{.18\textwidth}
	\centering
		\includegraphics[scale = 0.33]{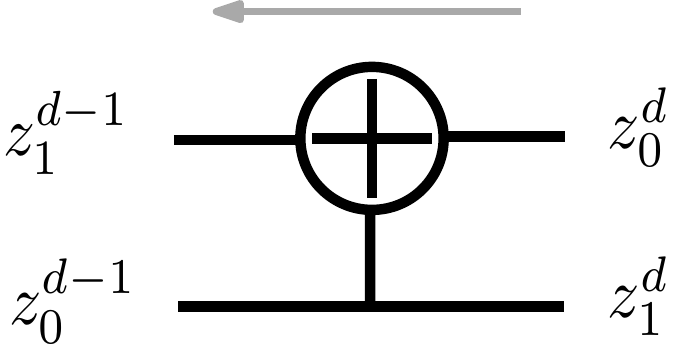}
		\caption{Sum junction}
		\label{fig:const_n7:junction}
	\end{subfigure}
	\begin{subfigure}[]{.3\textwidth}
	\centering
		\includegraphics[scale = 0.25]{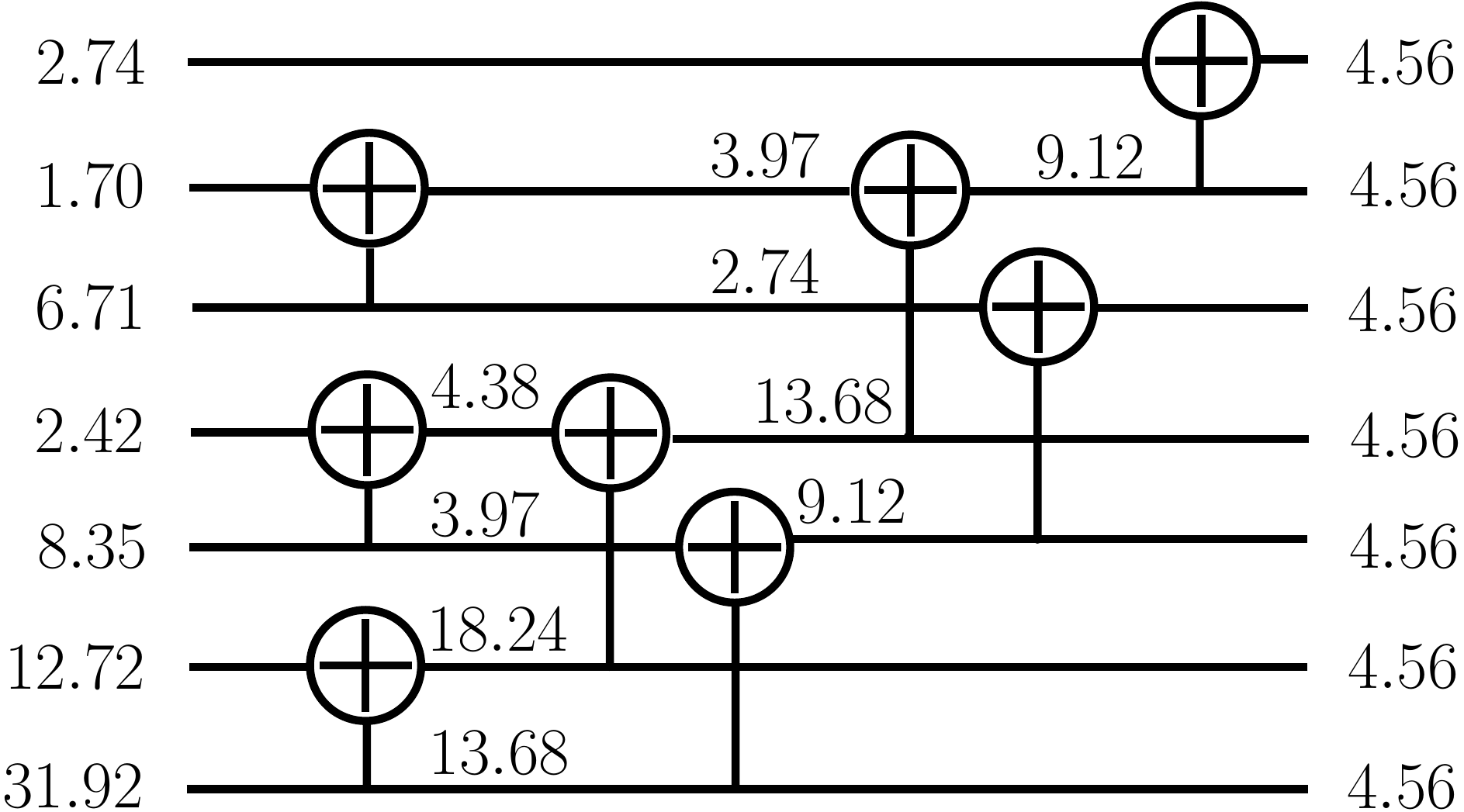}
		\caption{GA for $N_A=7$}
		\vspace{-.3\baselineskip}
		\label{fig:const_n7:ga}
	\end{subfigure}
	}
	\caption{Gaussian approximation reliability ordering of an ascending asymmetric polar code of length $N_A=7$ with $\bm{N} = \{4,2,1\}$.}
	\vspace{-\baselineskip}
\end{figure}

Most code construction algorithms used to build reliability sets for Ar{\i}kan codes can be adapted to APCs
with minor adjustments. Gaussian approximation \cite{Trifonov2012} (GA) serves as an effective 
method for designing frozen sets. One must consider the asymmetry of the Tanner graphs of APCs in order to
carry out the algorithm accurately. GA entails tracking the reliability of each bit at each stage of the encoder, where the reliabilities
are represented by the mean of the LLRs of the synthetic channel.
To begin, each coded bit in the encoder is assigned the LLR mean of the uncoded channel $z= 4R\mathcal{M}\frac{E_b}{N_0}$, where $\mathcal{M}$ is the modulation order.
The reliabilities are transformed at each summation junction between stage $d$ and $d-1$, as seen in Fig. \ref{fig:const_n7:junction}, according to
\begin{align}
	\label{eq:GA}
	 z_{0}^{d-1} &= \phi^{-1}(1-(1-\phi(z_0^{d} ))(1-\phi(z_1^{d} ))), \\
	 z_{1}^{d-1} &= z_0^{d} + z_1^{d}, \nonumber
\end{align}
where $\phi(x)$ and $\phi^{-1}(x)$ can be Trifinov's exact formulas \cite{Trifonov2012}, or approximations from the open source
simulation tool \textit{aff3ct} \cite{Cassagne2017a}. 
Eq. \ref{eq:GA} differs from the original GA method in that it cannot assume that each junction input has been transformed equally.
Computing the entire graph results in vector of subchannel reliabilities, which serves as a basis to rank indices 
and form $\mathcal{R}$.
A construction example for $\mathcal{PC}(7,4)$ with $\bm{N} = \{4,2,1\}$ is now presented for a target $\frac{E_b}{N_0}=3\text{dB}$ where $W$ is an AWGN channel using BPSK modulation. 
The full propagation is detailed in Fig. \ref{fig:const_n7:ga}. In this case, $\mathcal{R}=\{6,5,4,2,0,3,1\}$,
$\mathcal{I}=\{6,5,4,2\}$, and $\mathcal{F}=\{0,3,1\}$.

\subsection{Decoding}
\label{sec:apc:decode}
Just as when decoding Ar{\i}kan codes, the operations $f$ and $g$ from eq. (\ref{eq:f_g}) are used for decoding APCs since the 
same polarizing transform is used. As such, all standard polar decoders can be used with only a change in schedule
according to the new Tanner graph. The SC decoding tree has $p$ partial code trees that are children of asymmetric nodes and are 
decoded in the same manner as described in Section \ref{sec:pc:decode}.
The $p-1$ asymmetric nodes are decoded with a schedule that corresponds to the additional polarizing stages outlined in Section \ref{sec:apc:gen_matrix}.
Specifically, each sum junction corresponds to an $f$ and $g$ function with equivalent indexing. 

The number of computations required by an asymmetric
SC decoder is at most $\sum_{l=0}^{p-1} N_l\log_2{N_l} \!+\! \sum_{l=1}^{p\!-\!1} 2N_l$, which is proven in the Appendix to be always less than 
equivalent PS codes, namely $N_M \log_2N_M$ where $N_M = \lceil \log_2N_A \rceil$. In the case where $p\le2$, the ascending and descending APCs have the same number
of decoding operations. Generally, the descending permutation requires a higher number of operations.  

\begin{figure}[t]
\vspace*{.3\baselineskip}
\makebox[\linewidth][c]{
	\centering
	\hspace{-6mm}
	\begin{subfigure}[t]{.235\textwidth}
	\centering
		\input{figs/n14_binary_tree_modified.tikz}
		\vspace*{-1.5\baselineskip}
		\caption{SC}
	\end{subfigure}
	\begin{subfigure}[t]{.23\textwidth}
	\centering
		\input{figs/n14_binary_tree_fssc_modified.tikz}
		\vspace*{-1.5\baselineskip}
		\caption{Fast-SSC}
	\end{subfigure}
	}
	\caption{SC tree for an ascending asymmetric polar code with $N_A=14$ and $\bm{N} = \{8,4,2\}$. White, blue, and red leaves are frozen bits, Rate-0 nodes, and Rate-1 nodes, respectively. Subscripts indicate node or function size.}
	\label{fig:n14_tree}
\end{figure}
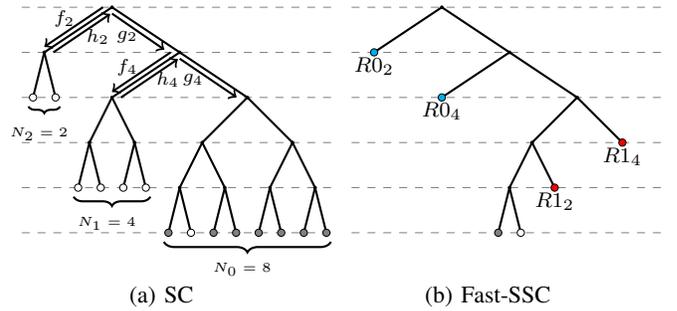

Fast-SSC decoding schedules can be obtained for APCs since all symmetric nodes remain powers of 2. The SC tree for an ascending APC where $N_A = 14$, along with its Fast-SSC counterpart, can be
seen in Fig. \ref{fig:n14_tree}.
Regarding error correction performance, the theoretical FER under SC decoding for APCs can be 
analytically computed using GA bit reliabilities, as was observed for Ar{\i}kan polar codes in \cite[eq. (3)]{Wu2014}. 

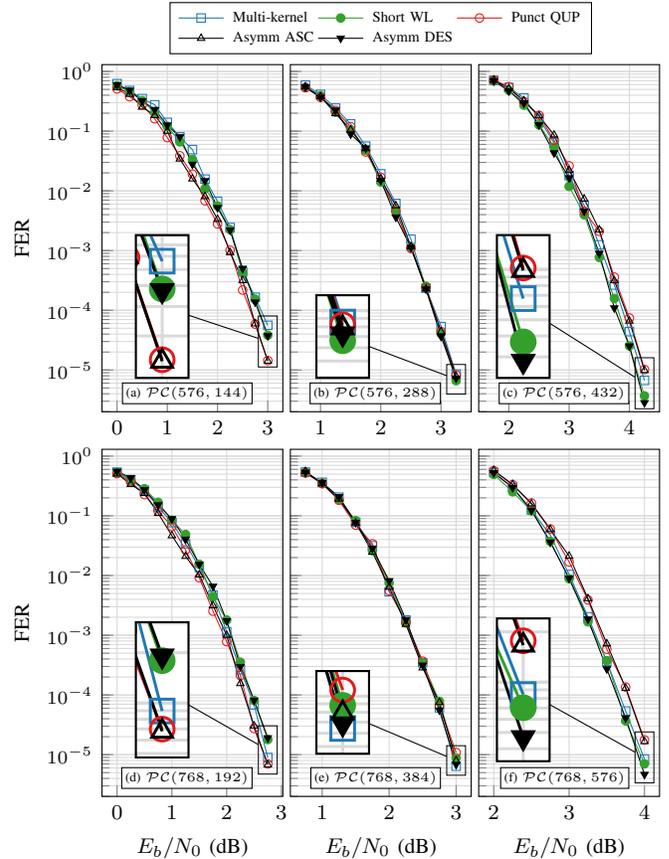
\begin{figure}[t]
	\vspace{-.5\baselineskip}
	\hspace{5mm}
	\centering
		\input{figs/big_576_768_group_fer_plot.tikz}
		\vspace{-1.3\baselineskip}
	\caption{FER curves for polar codes with $N\in\{576, 768\}$ (top to bottom) with rates $R\in\{\frac{1}{4},\frac{1}{2},\frac{3}{4}\}$ (left to right).}
	\vspace{-\baselineskip}
	\label{fig:fer:576_768}
\end{figure}

\section{Numerical Illustrations and Analysis}
\label{sec:numerical_analysis}
\subsection{Decoding Performance}
\label{sec:numerical_analysis:performance}
The error correction capabilities of APCs has been evaluated through a series of simulations using the AWGN channel and BPSK
modulation. FER curves have been obtained for $N\in\{576, 768, 1536, 2304, 3072\}$ and $R\in\{\frac{1}{4}, \frac{1}{2},
 \frac{3}{4}\}$. SCL-CRC is the decoding algorithm used with list size $L=8$ and CRC size 16 using polynomial
$0 \hspace{-.6mm} \times \hspace{-1mm} 1021$. All frozen sets were reconstructed for each $\frac{E_b}{N_0}$ value in the plots 
using GA. We compared APCs with two key PS schemes as well as with MK polar codes. The kernel 
order of the MK codes was optimized for highest overall reliability by an exhaustive search, as proposed in \cite{Gabry2016}.

\begin{table}
\vspace{-\baselineskip}
	\centering
	\begin{tabular}{@{}llllll@{}}
	\toprule
	Scheme 	& $N\!=\!576$	& $N\!=\!768$  	& $N\!=\!1536$	& $N\!=\!2304$	& $N\!=\!3072$	\\
	\cmidrule(r){1-1}
	\cmidrule(l){2-2}
	\cmidrule(l){3-3}
	\cmidrule(l){4-4}
	\cmidrule(l){5-5}
	\cmidrule(l){6-6}
	APC    & 5120       & 7168 		& 15872 	& 25088 	& 34816 	\\
	PS     & 10240  	& 10240 	& 22528 	& 49152 	& 49152     \\
	MK     & 4608   	& 6912 		& 15360 	& 23040 	& 33792     \\
	\bottomrule
	\end{tabular}
		\caption{Number of SC decoding operations.}
	\vspace{-\baselineskip}
	\label{table:num_ops}
\end{table}

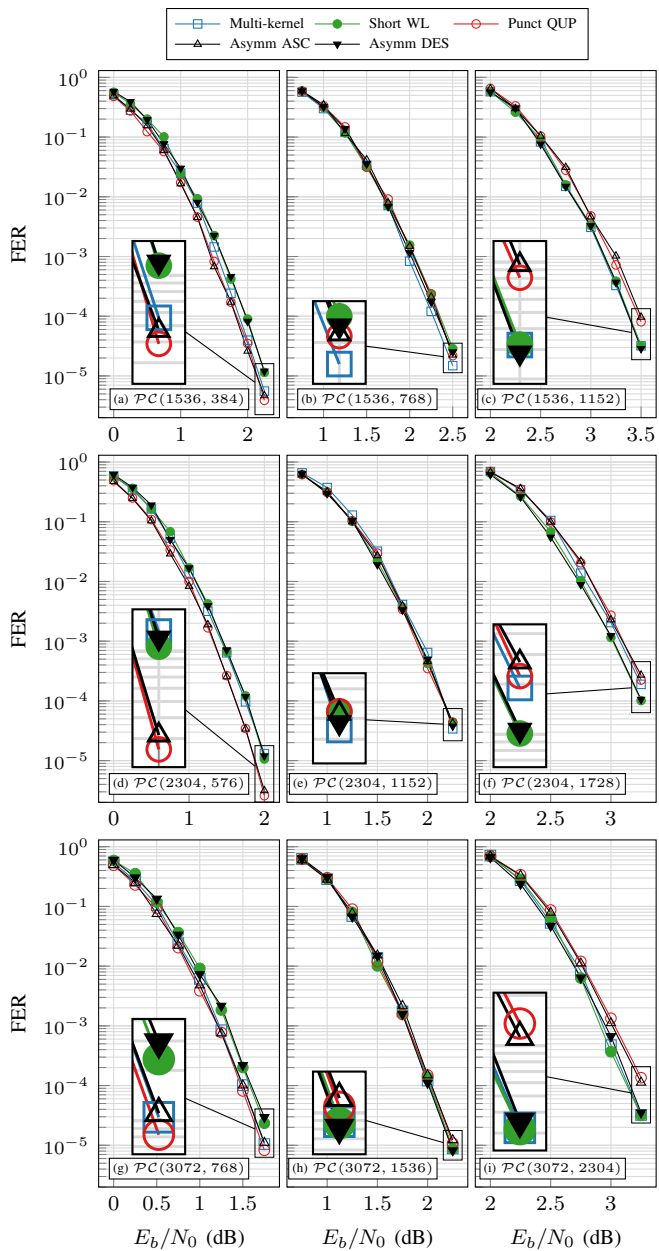
\begin{figure}[t]
	\hspace{-4mm}
	\centering
		\input{figs/big_1536_2304_3072_group_fer_plot.tikz}
		\vspace{-.4\baselineskip}
	\caption{FER curves for polar codes with $N\in\{1536, 2304, 3072\}$ (top to bottom) with rates $R\in\{\frac{1}{4},\frac{1}{2},\frac{3}{4}\}$ (left to right).}
	\vspace{-\baselineskip}
	\label{fig:fer:1536_2304_3072}
\end{figure}

Under SCL-CRC decoding, asymmetric codes have comparable performance to leading length-compatible schemes when
using a range of code rates, as depicted in Figs. \ref{fig:fer:576_768} and \ref{fig:fer:1536_2304_3072}. 
In all cases considered, APC performance is in the approximate range of $[-0.05\text{dB},+0.05\text{dB}]$ compared to the 
best performing state-of-the-art codes for $\text{FER}=10^{-4}$.
Just as with Ar{\i}kan polar codes, APC error correction performance generally improves with length. It should be noted 
that APCs excel when they contain fewer and larger partial polar codes.

For all code lengths, APCs built with the ascending partial code permutation have superior error correction performance to APCs utilizing the descending permutation at low rates,
while the opposite is true for high rates. This is due to the fact that the reliability of partial codes is related not only to their location in the Tanner graph, but also their
size. Further, APCs have reduced performance when indices in $\mathcal{I}$ are found in smaller partial codes. The small partial codes in the descending permutation are where 
the few highly reliable indices are located at low rates, and so these bits exhibit reduced polarization effects due to the smaller partial code size. 
Conversely, the ascending permutation allows most information bits to be located in the largest partial codes at low rates, which present the highest reliability. 
It should also be noted that ascending and descending APCs have very similar performance to punctured and shortened polar codes, respectively. MK codes generally 
do not outperform PS or APC schemes in the tested scenarios. However, they are more likely to have worse performance when they have more than one ternary stage,
as seen in Figs. \ref{fig:fer:576_768}ac and \ref{fig:fer:1536_2304_3072}df. 

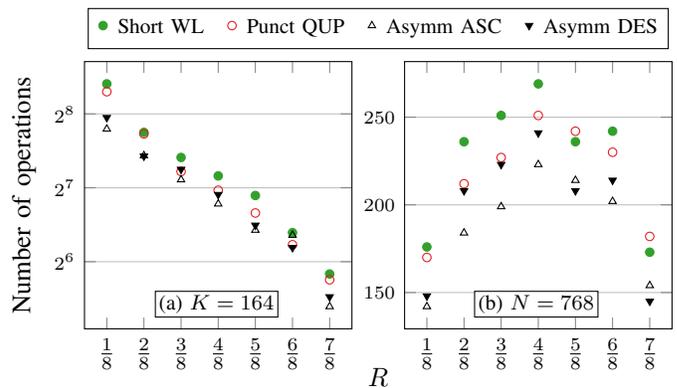
\begin{figure}[t]
	\hspace{-2mm}
	\centering
			\input{figs/fssc_ops_group_plot.tikz}
			\vspace{-\baselineskip}
	\caption{Fast-SSC complexity with a maximum processor size of 64.}
	\vspace{-\baselineskip}
	\label{fig:fssc_plot}
\end{figure}

\subsection{Complexity Evaluation}
\label{sec:numerical_analysis:complexity}
We will analyze the complexity of APCs by comparing them with PS and MK schemes in terms of SC and Fast-SSC decoding complexity,
space complexity, and code design.
We measured the SC time complexity of all considered techniques by examining the number of LLR operations
required for decoding a single codeword. The number of operations required for each scheme is given in Sections \ref{sec:pc:decode}, 
\ref{sec:pc:mk}, and \ref{sec:apc:decode}. Table \ref{table:num_ops} outlines the number of decoding operations required for each simulated case.
Observe that both APC and MK codes have comparable decoding complexity that is directly related to their block length.
Although MK codes have reduced time complexity
over equivalent PS and APC codes, MK decoding requires concessions for ternary decoding using eq. (\ref{eq:mk_dec}), 
and kernel order in scheduling. The logic required for practical MK implementations
is more complex and demands specialized hardware and increased memory over Ar{\i}kan codes \cite{Coppolino2018}.
Further, MK codes are limited in their length flexibility without considering the use of higher order kernels, which introduces further complication.
As such, MK complexity is difficult to compare with that of APCs or PS codes.

Comparing Fast-SSC decoding complexity requires a more refined metric. Since there are no fast decoders currently available for MK codes,
they will be excluded from this comparison. It was shown in \cite{Leroux2013} that a processing element size of 64
permits an acceptable balance between decoder throughput and hardware utilization for Ar{\i}kan polar codes with block length $N=1024$. 
We will use this baseline for our comparisons, seeing that our experiments operate on a similar order of magnitude. 
Given that specialized nodes such as Rate-1, Rate-0, Single-Parity Check, and Repetition \cite{Sarkis2014}
are available with a maximum node size of 64, each specialized node is considered a single operation. 
Additionally, all computations of $\bm{\alpha^l}$ and $\bm{\alpha^r}$ of children nodes are counted as $\lceil\frac{N_v}{64}\rceil$ operation(s), 
where $N_v$ is the child node size. It is worth noting that PS decoders are likely to have a higher number of Rate-0 and Repetition nodes due to their
higher proportion of frozen bits. However, Fig. \ref{fig:fssc_plot} demonstrates that APCs have between $[2.7\%,27\%]$ 
time complexity reduction under Fast-SSC decoding when compared against PS schemes over a range of block lengths and rates. The degree to which APC complexity 
is reduced is dependent on the difference between the transmitted block length and the size of the PS mother code. When $N$ is slightly larger than a power of 2, 
APCs exhibit higher complexity reduction.

Regarding space complexity, APCs have the smaller memory requirements than PS schemes. Using the efficient SC implementations proposed in \cite{Leroux2013}, 
APCs only require $\bm{\alpha}$ and $\bm{\beta}$ memory for the largest partial code, which is at most half that of equivalent PS codes. For example, 
when $N=2304$, PS decoders require the same $\bm{\alpha}$ and $\bm{\beta}$ storage capacity as an Ar{\i}kan code of $N=4096$, while APCs can be decoded
with just the memory capacity of a $N=2048$ Ar{\i}kan code and extended channel memory. Conclusively, PS approaches 
are not an efficient use of time and space resources since they must decode their mother code in order to receive an effectively smaller codeword. 
Further, MK codes necessitate ternary $\bm{\beta}$ memory, which results in increased
storage requirements over purely Ar{\i}kan decoders \cite{Coppolino2018}.

Regarding the code construction, 
MK schemes must optimize the kernel order to maximize performance, and PS schemes
must build a reliability set after the PS sets are determined, or vice versa. 
Thus, APCs demand only a single-step frozen set construction, while the competing schemes call for further considerations. 

\section{Conclusion}
\label{sec:conclusion}
In this work, we introduced the concept of \textit{asymmetric polar codes} and presented techniques to retain known decoders
and construction methods. We analyzed our APC construction and evaluated its utility in terms of FER performance and
decoding complexity. We compared APCs with state of the art PS and MK codes through simulation.
We have demonstrated that APCs are a flexible polar coding scheme that exhibit block length-dependent decoding complexity with similar error
correction performance to the state-of-the-art. We established that APCs are a practical family to be added to the polar code tool box. 
Future work in APCs will involve flexible HARQ retransmission schemes and low complexity frozen set design. 

\section*{Appendix}

\subsubsection*{Proof of reduced decoding complexity} 
\label{sec:appendix:proof}
We will show that APCs always require fewer computations than an
equivalent PS code by examining the worst case scenario: when $asc=0$ and $N = 2^n - 1$. 
This case yields the highest possible $p$. If
\vspace{-.5\baselineskip}
\begin{equation*}
\sum_{l=0}^{p-1} N_l\log_2{N_l} + \sum_{l=1}^{p-1} 2N_l < 2^{\lceil \log_2N \rceil} \log_2{2^{\lceil \log_2N \rceil}}, 
\end{equation*}
is true in the worst case, then APCs always require fewer decoding operations than an equivalent PS code for any $N$. In this case,
$p = \lceil \log_2N \rceil = \log_2{(N+1)}$ and $\bm{N}$ contains all powers
of two less than $N_M$:
\vspace{-.4\baselineskip}

\small
\begin{equation*}
\begin{aligned}
\sum_{l=0}^{p-1} 2^l\log_2{2^l} + \sum_{l=1}^{p-1} 2\cdot2^{l} &< 2^p \log_2{2^p}, \\
\sum_{l=0}^{p-1} l \cdot2^l + \sum_{l=1}^{p-1} 2\cdot2^l &< p \cdot2^p, \\
p \cdot2^p -2 &< p \cdot2^p. \\
\end{aligned}
\end{equation*}
\normalsize
Thus, APCs have lower decoding complexity than PS codes. 
It can be shown that APCs have fewer decoding operations when $asc=1$ than when $asc=0$ by replacing the second summation 
term with $\sum_{l=1}^{p-1} 2\cdot2^{l-1}$, which results in a maximum number of decoding operations of $p \cdot2^p -2^p < p \cdot2^p$.

\bibliographystyle{IEEEtran}
\bibliography{IEEEabrv,asymmetric}

\end{document}

%% file: figs/n14_binary_tree_modified.tikz
\begin{tikzpicture}[scale=1.2, thick]

\draw [very thin,gray,dashed] (-2,.5) -- (1.5,.5);
\draw [very thin,gray,dashed] (-2,0) -- (1.5,0);
\draw [very thin,gray,dashed] (-2,-.5) -- (1.5,-.5);
\draw [very thin,gray,dashed] (-2,-1) -- (1.5,-1);
\draw [very thin,gray,dashed] (-2,-1.5) -- (1.5,-1.5);
\draw [very thin,gray,dashed] (-2,-2) -- (1.5,-2);

\fill (-1,.5) circle [radius=.02];

\fill (-.25,0) circle [radius=.02];
\fill (-1.75,0) circle [radius=.02];

\fill (-1,-1+.5) circle [radius=.02];
\fill (.5,-1+.5) circle [radius=.02];

\fill (-1.25,-1.5+.5) circle [radius=.02];
\fill (-.75,-1.5+.5) circle [radius=.02];

\fill (-1+1,-1) circle [radius=.02];
\fill (-1+2,-1) circle [radius=.02];

\fill (-1.25+1,-1.5) circle [radius=.02];
\fill (-.75+1,-1.5) circle [radius=.02];
\fill (-1.25+2,-1.5) circle [radius=.02];
\fill (-.75+2,-1.5) circle [radius=.02];


\draw [->] (-1.1,.5) -- (-1.75,0.062);
\draw [<-] (-1.02,.425) -- (-1.65,-.00);
\draw [->] (-1,.43) -- (-.42,0.03);
\node at (-1.525,0.375) {\scriptsize $f_2$};
\node at (-1.155,0.175) {\scriptsize $h_2$};
\node at (-.83,0.175) {\scriptsize $g_2$};

\draw [->] (-.35,0) -- (-1,-.438);
\draw [<-] (-.27,-0.075) -- (-0.9,-.5);
\draw [->] (-.25,-.07) -- (.38,-.497);
\node at (-.83,-.16) {\scriptsize $f_4$};
\node at (-.38,-.3) {\scriptsize $h_4$};
\node at (-.1,-.3) {\scriptsize $g_4$};


\draw [black,decorate,decoration={brace,amplitude=4pt}] (1.425,-2.1)  -- (-.425,-2.1) 
	node [black,midway,below=4pt,xshift=-2pt] {\tiny $N_0 = 8$};
\draw [black,decorate,decoration={brace,amplitude=3.5pt}] (-.575,-1.6)  -- (-1.425,-1.6) 
	node [black,midway,below=4pt,xshift=-2pt] {\tiny $N_1 = 4$};
\draw [black,decorate,decoration={brace,amplitude=2.5pt}] (-1.575,-.6)  -- (-1.925,-.6) 
	node [black,midway,below=4pt,xshift=-2pt] {\tiny $N_2 = 2$};


\draw (-1,.5) -- (-1.75,0);
\draw (-1,.5) -- (-.25,0);

\draw (-.25,0) -- (-1,-1+.5);
\draw (-.25,0) -- (.5,-1+.5);

\draw (-1.375+1.875,-.5) -- (-1.75+1.75,-1);
\draw (-1.375+1.875,-.5) -- (-1+2,-1);
\draw (-1.75,0) -- (-1.875,-.5);
\draw (-1.75,0) -- (-1.625,-.5);

\draw (-1,-1.0+.5) -- (-1.25,-1.5+.5);
\draw (-1,-1.0+.5) -- (-.75,-1.5+.5);

\draw (-1+1,-1.0) -- (-1.25+1,-1.5);
\draw (-1+1,-1.0) -- (-.75+1,-1.5);
\draw (-1+2,-1.0) -- (-1.25+2,-1.5);
\draw (-1+2,-1.0) -- (-.75+2,-1.5);

\draw (-1.25,-1.5+.5) -- (-1.375,-2+.5);
\draw (-1.25,-1.5+.5) -- (-1.125,-2+.5);
\draw (-.75,-1.5+.5) -- (-.875,-2+.5);
\draw (-.75,-1.5+.5) -- (-.625,-2+.5);

\draw (-1.25+1,-1.5) -- (-1.375+1,-2);
\draw (-1.25+1,-1.5) -- (-1.125+1,-2);
\draw (-.75+1,-1.5) -- (-.875+1,-2);
\draw (-.75+1,-1.5) -- (-.625+1,-2);
\draw (-1.25+2,-1.5) -- (-1.375+2,-2);
\draw (-1.25+2,-1.5) -- (-1.125+2,-2);
\draw (-.75+2,-1.5) -- (-.875+2,-2);
\draw (-.75+2,-1.5) -- (-.625+2,-2);

\fill [white](-1.875,-.5) circle [radius=.04];
\fill [white](-1.625,-.5) circle [radius=.04];
\fill [white](-1.375,-2+.5) circle [radius=.04];
\fill [white](-1.125,-2+.5) circle [radius=.04];
\fill [white](-.875,-2+.5) circle [radius=.04];
\fill [white](-.625,-2+.5) circle [radius=.04];
\fill [gray](-.375,-2) circle [radius=.04];
\fill [white](-.125,-2) circle [radius=.04];
\fill [gray](.125,-2) circle [radius=.04];
\fill [gray](.375,-2) circle [radius=.04];
\fill [gray](.625,-2) circle [radius=.04];
\fill [gray](.875,-2) circle [radius=.04];
\fill [gray](1.125,-2) circle [radius=.04];
\fill [gray](1.375,-2) circle [radius=.04];

\draw [very thin] (-1.875,-.5) circle [radius=.04];
\draw [very thin] (-1.625,-.5) circle [radius=.04];
\draw [very thin] (-1.375,-2+.5) circle [radius=.04];
\draw [very thin] (-1.125,-2+.5) circle [radius=.04];
\draw [very thin] (-.875,-2+.5) circle [radius=.04];
\draw [very thin] (-.625,-2+.5) circle [radius=.04];
\draw [very thin] (-.375,-2) circle [radius=.04];
\draw [very thin] (-.125,-2) circle [radius=.04];
\draw [very thin] (.125,-2) circle [radius=.04];
\draw [very thin] (.375,-2) circle [radius=.04];
\draw [very thin] (.625,-2) circle [radius=.04];
\draw [very thin] (.875,-2) circle [radius=.04];
\draw [very thin] (1.125,-2) circle [radius=.04];
\draw [very thin] (1.375,-2) circle [radius=.04];


\end{tikzpicture}

%% file: figs/n14_binary_tree_fssc_modified.tikz
\begin{tikzpicture}[scale=1.2, thick]

\draw [very thin,gray,dashed] (-2,.5) -- (1.5,.5);
\draw [very thin,gray,dashed] (-2,0) -- (1.5,0);
\draw [very thin,gray,dashed] (-2,-.5) -- (1.5,-.5);
\draw [very thin,gray,dashed] (-2,-1) -- (1.5,-1);
\draw [very thin,gray,dashed] (-2,-1.5) -- (1.5,-1.5);
\draw [very thin,gray,dashed] (-2,-2) -- (1.5,-2);

\fill (-1,.5) circle [radius=.02];

\fill (-.25,0) circle [radius=.02];

\fill (.5,-1+.5) circle [radius=.02];

\fill (-1+1,-1) circle [radius=.02];

\fill (-1.25+1,-1.5) circle [radius=.02];
\fill (-.75+1,-1.5) circle [radius=.02];


\draw [white,decorate,decoration={brace,amplitude=4pt}] (1.425,-2.1)  -- (-.425,-2.1) 
	node [white,midway,below=4pt,xshift=-2pt] {\tiny $N_0 = 8$};
\draw [white,decorate,decoration={brace,amplitude=3.5pt}] (-.575,-1.6)  -- (-1.425,-1.6) 
	node [white,midway,below=4pt,xshift=-2pt] {\tiny $N_1 = 4$};
\draw [white,decorate,decoration={brace,amplitude=2.5pt}] (-1.575,-.6)  -- (-1.925,-.6) 
	node [white,midway,below=4pt,xshift=-2pt] {\tiny $N_2 = 2$};


\draw (-1,.5) -- (-1.75,0);
\draw (-1,.5) -- (-.25,0);

\draw (-.25,0) -- (-1,-1+.5);
\draw (-.25,0) -- (.5,-1+.5);

\draw (-1.375+1.875,-.5) -- (-1.75+1.75,-1);
\draw (-1.375+1.875,-.5) -- (-1+2,-1);

\draw (-1+1,-1.0) -- (-1.25+1,-1.5);
\draw (-1+1,-1.0) -- (-.75+1,-1.5);

\draw (-1.25+1,-1.5) -- (-1.375+1,-2);
\draw (-1.25+1,-1.5) -- (-1.125+1,-2);

\fill [cyan](-1.75,0) circle [radius=.04];
\draw [very thin] (-1.75,0) circle [radius=.04];

\fill [cyan](-1,-1+.5) circle [radius=.04];
\draw [very thin] (-1,-1+.5) circle [radius=.04];

\fill [red](-.75+1,-1.5) circle [radius=.04];
\draw [very thin] (-.75+1,-1.5) circle [radius=.04];

\fill [red](-1+2,-1) circle [radius=.04];
\draw [very thin] (-1+2,-1) circle [radius=.04];

\fill [gray](-.375,-2) circle [radius=.04];
\fill [white](-.125,-2) circle [radius=.04];

\draw [very thin] (-.375,-2) circle [radius=.04];
\draw [very thin] (-.125,-2) circle [radius=.04];


\node at (-1.75,-.15) {\footnotesize $R0_2$};
\node at (-1,-1+.35) {\footnotesize $R0_4$};
\node at (-.75+1,-1.65) {\footnotesize $R1_2$};
\node at (-1+2,-1.15) {\footnotesize $R1_4$};

\end{tikzpicture}

%% file: figs/big_576_768_group_fer_plot.tikz
\begin{tikzpicture}
[spy using outlines={chamfered rectangle, magnification=2.8, connect spies}]

\begin{groupplot}[
        group style={group name=fer_group, group size= 3 by 6, horizontal sep=1mm, vertical sep=5mm,}, height=0.7\linewidth, width=.45\linewidth]


        \nextgroupplot[ylabel=FER, grid=both, grid style={gray!30}, tick align=inside, tickpos=left, yticklabel style = {font=\scriptsize}, ylabel style = {font=\footnotesize},  xticklabel style = {font=\footnotesize}, ymode=log, ymax=1.3,  ymin=2e-6]
                \begin{scope}
                        \addplot[Paired-1, mark=square, mark size=1.6pt, thin] table [x={EbN0}, y={FER}, col sep=comma] {figs/data/576_144/576,144_mk_h.csv};
                        \addplot[Paired-3, mark=*, mark size=1.6pt, thin] table [x={EbN0}, y={FER}, col sep=comma] {figs/data/576_144/576,144_s_w.csv};
                        \addplot[Paired-5, mark=o, mark size=1.6pt, thin] table [x={EbN0}, y={FER}, col sep=comma] {figs/data/576_144/576,144_p_q.csv};
                        \addplot[black, mark=triangle, mark size=1.6pt, thin] table [x={EbN0}, y={FER}, col sep=comma] {figs/data/576_144/576,144_apc_a.csv};
                        \addplot[black, mark=triangle*, mark options={rotate=180}, mark size=1.6pt, thin] table [x={EbN0}, y={FER}, col sep=comma] {figs/data/576_144/576,144_apc_d.csv};

                        \coordinate (zoom1) at (axis cs:3,3e-5);
                        \spy [black, height=1.9cm, width=0.7cm] on (zoom1) in node [fill=white] at (0.8,1.4);
                \end{scope}

        \nextgroupplot[grid=both, grid style={gray!30}, tick align=inside, tickpos=left, yticklabels={}, xticklabel style = {font=\footnotesize}, ymode=log, ymax=1.3,  ymin=2e-6,
                        legend style={font=\footnotesize, nodes={scale=0.7}, at={(0.5,1.03)}, anchor=south, legend columns=3},
                        legend cell align={left},
                        legend entries={~{Multi-kernel}, ~{Short WL}, ~{Punct QUP}, ~{Asymm ASC}, ~{Asymm DES}}
                        ]
                \begin{scope}
                        \addplot[Paired-1, mark=square, mark size=1.6pt, thin] table [x={EbN0}, y={FER}, col sep=comma] {figs/data/576_288/576,288_mk_h.csv};
                        \addplot[Paired-3, mark=*, mark size=1.6pt, thin] table [x={EbN0}, y={FER}, col sep=comma] {figs/data/576_288/576,288_s_w.csv};
                        \addplot[Paired-5, mark=o, mark size=1.6pt, thin] table [x={EbN0}, y={FER}, col sep=comma] {figs/data/576_288/576,288_p_q.csv};
                        \addplot[black, mark=triangle, mark size=1.6pt, thin] table [x={EbN0}, y={FER}, col sep=comma] {figs/data/576_288/576,288_apc_a.csv};
                        \addplot[black, mark=triangle*, mark options={rotate=180}, mark size=1.6pt, thin] table [x={EbN0}, y={FER}, col sep=comma] {figs/data/576_288/576,288_apc_d.csv};

                        \coordinate (zoom2) at (axis cs:3.25,7e-6);
                        \spy [black, height=1.1cm, width=0.7cm] on (zoom2) in node [fill=white] at (3.2,1);
                \end{scope}

        \nextgroupplot[grid=both, grid style={gray!30}, tick align=inside, tickpos=left, yticklabels={}, ylabel style = {font=\footnotesize}, xticklabel style = {font=\footnotesize}, ymode=log, ymax=1.3,  ymin=2e-6]
                \begin{scope}
                        \addplot[Paired-1, mark=square, mark size=1.6pt, thin] table [x={EbN0}, y={FER}, col sep=comma] {figs/data/576_432/576,432_mk_h.csv};
                        \addplot[Paired-3, mark=*, mark size=1.6pt, thin] table [x={EbN0}, y={FER}, col sep=comma] {figs/data/576_432/576,432_s_w.csv};
                        \addplot[Paired-5, mark=o, mark size=1.6pt, thin] table [x={EbN0}, y={FER}, col sep=comma] {figs/data/576_432/576,432_p_q.csv};
                        \addplot[black, mark=triangle, mark size=1.6pt, thin] table [x={EbN0}, y={FER}, col sep=comma] {figs/data/576_432/576,432_apc_a.csv};
                        \addplot[black, mark=triangle*, mark options={rotate=180}, mark size=1.6pt, thin] table [x={EbN0}, y={FER}, col sep=comma] {figs/data/576_432/576,432_apc_d.csv};

                        \coordinate (zoom3) at (axis cs:4.25,6e-6);
                        \spy [black, height=1.9cm, width=0.7cm] on (zoom3) in node [fill=white] at (5.6,1.4);
                \end{scope}


        \nextgroupplot[xlabel= \footnotesize {${E_b/N_0}$ (dB)}, ylabel=FER, grid=both, grid style={gray!30}, tick align=inside, tickpos=left, yticklabel style = {font=\scriptsize}, ylabel style = {font=\footnotesize},  xticklabel style = {font=\footnotesize}, ymode=log, ymax=1.3,  ymin=2e-6]
                \begin{scope}
                        \addplot[Paired-1, mark=square, mark size=1.6pt, thin] table [x={EbN0}, y={FER}, col sep=comma] {figs/data/768_192/768,192_mk_h.csv};
                        \addplot[Paired-3, mark=*, mark size=1.6pt, thin] table [x={EbN0}, y={FER}, col sep=comma] {figs/data/768_192/768,192_s_w.csv};
                        \addplot[Paired-5, mark=o, mark size=1.6pt, thin] table [x={EbN0}, y={FER}, col sep=comma] {figs/data/768_192/768,192_p_q.csv};
                        \addplot[black, mark=triangle, mark size=1.6pt, thin] table [x={EbN0}, y={FER}, col sep=comma] {figs/data/768_192/768,192_apc_a.csv};
                        \addplot[black, mark=triangle*, mark options={rotate=180}, mark size=1.6pt, thin] table [x={EbN0}, y={FER}, col sep=comma] {figs/data/768_192/768,192_apc_d.csv};

                        \coordinate (zoom4) at (axis cs:2.75,1.2e-5);
                        \spy [black, height=1.8cm, width=0.7cm] on (zoom4) in node [fill=white] at (0.8,-3.7);
                \end{scope}

        \nextgroupplot[xlabel= \footnotesize {${E_b/N_0}$ (dB)}, grid=both, grid style={gray!30}, tick align=inside, tickpos=left, yticklabels={}, ylabel style = {font=\footnotesize}, xticklabel style = {font=\footnotesize}, ymode=log, ymax=1.3,  ymin=2e-6]
                \begin{scope}
                        \addplot[Paired-1, mark=square, mark size=1.6pt, thin] table [x={EbN0}, y={FER}, col sep=comma] {figs/data/768_384/768,384_mk_h.csv};
                        \addplot[Paired-3, mark=*, mark size=1.6pt, thin] table [x={EbN0}, y={FER}, col sep=comma] {figs/data/768_384/768,384_s_w.csv};
                        \addplot[Paired-5, mark=o, mark size=1.6pt, thin] table [x={EbN0}, y={FER}, col sep=comma] {figs/data/768_384/768,384_p_q.csv};
                        \addplot[black, mark=triangle, mark size=1.6pt, thin] table [x={EbN0}, y={FER}, col sep=comma] {figs/data/768_384/768,384_apc_a.csv};
                        \addplot[black, mark=triangle*, mark options={rotate=180}, mark size=1.6pt, thin] table [x={EbN0}, y={FER}, col sep=comma] {figs/data/768_384/768,384_apc_d.csv};

                        \coordinate (zoom5) at (axis cs:3,8e-6);
                        \spy [black, height=1.1cm, width=0.7cm] on (zoom5) in node [fill=white] at (3.2,-4);
                \end{scope}

        \nextgroupplot[xlabel= \footnotesize {${E_b/N_0}$ (dB)}, grid=both, grid style={gray!30}, tick align=inside, tickpos=left, yticklabels={}, ylabel style = {font=\footnotesize}, xticklabel style = {font=\footnotesize}, ymode=log, ymax=1.3,  ymin=2e-6]
                \begin{scope}
                        \addplot[Paired-1, mark=square, mark size=1.6pt, thin] table [x={EbN0}, y={FER}, col sep=comma] {figs/data/768_576/768,576_mk_h.csv};
                        \addplot[Paired-3, mark=*, mark size=1.6pt, thin] table [x={EbN0}, y={FER}, col sep=comma] {figs/data/768_576/768,576_s_w.csv};
                        \addplot[Paired-5, mark=o, mark size=1.6pt, thin] table [x={EbN0}, y={FER}, col sep=comma] {figs/data/768_576/768,576_p_q.csv};
                        \addplot[black, mark=triangle, mark size=1.6pt, thin] table [x={EbN0}, y={FER}, col sep=comma] {figs/data/768_576/768,576_apc_a.csv};
                        \addplot[black, mark=triangle*, mark options={rotate=180}, mark size=1.6pt, thin] table [x={EbN0}, y={FER}, col sep=comma] {figs/data/768_576/768,576_apc_d.csv};

                        \coordinate (zoom6) at (axis cs:4,9e-6);
                        \spy [black, height=1.9cm, width=0.7cm] on (zoom6) in node [fill=white] at (5.6,-3.7);
                \end{scope}

\end{groupplot}



\node[fill = white, draw=black, inner sep=0.5mm, above left = 0.1cm and -0.8cm of fer_group c1r1.south] {\tiny (a) $\mathcal{PC}(576,144)$}; 
\node[fill = white, draw=black, inner sep=0.5mm, above left = 0.1cm and -0.8cm of fer_group c2r1.south] {\tiny (b) $\mathcal{PC}(576,288)$};
\node[fill = white, draw=black, inner sep=0.5mm, above left = 0.1cm and -0.8cm of fer_group c3r1.south] {\tiny (c) $\mathcal{PC}(576,432)$};

\node[fill = white, draw=black, inner sep=0.5mm, above left = 0.1cm and -0.8cm of fer_group c1r2.south] {\tiny (d) $\mathcal{PC}(768,192)$};
\node[fill = white, draw=black, inner sep=0.5mm, above left = 0.1cm and -0.8cm of fer_group c2r2.south] {\tiny (e) $\mathcal{PC}(768,384)$};
\node[fill = white, draw=black, inner sep=0.5mm, above left = 0.1cm and -0.8cm of fer_group c3r2.south] {\tiny (f) $\mathcal{PC}(768,576)$};

\end{tikzpicture}

%% file: figs/big_1536_2304_3072_group_fer_plot.tikz
\begin{tikzpicture}
[spy using outlines={chamfered rectangle, magnification=2.8, height=1cm, width=0.5cm, connect spies}]

\begin{groupplot}[
        group style={group name=fer_group, group size= 3 by 9, horizontal sep=1mm, vertical sep=5mm,}, height=0.7\linewidth, width=.45\linewidth]


        \nextgroupplot[ylabel=FER, grid=both, grid style={gray!30}, tick align=inside, tickpos=left, yticklabel style = {font=\scriptsize}, ylabel style = {font=\footnotesize},  xticklabel style = {font=\footnotesize}, ymode=log, ymax=1.3,  ymin=2e-6]
                \addplot[Paired-1, mark=square, mark size=1.6pt, thin] table [x={EbN0}, y={FER}, col sep=comma] {figs/data/1536_384/1536,384_mk_h.csv};
                \addplot[Paired-3, mark=*, mark size=1.6pt, thin] table [x={EbN0}, y={FER}, col sep=comma] {figs/data/1536_384/1536,384_s_w.csv};
                \addplot[Paired-5, mark=o, mark size=1.6pt, thin] table [x={EbN0}, y={FER}, col sep=comma] {figs/data/1536_384/1536,384_p_q.csv};
                \addplot[black, mark=triangle, mark size=1.6pt, thin] table [x={EbN0}, y={FER}, col sep=comma] {figs/data/1536_384/1536,384_apc_a.csv};
                \addplot[black, mark=triangle*, mark options={rotate=180}, mark size=1.6pt, thin] table [x={EbN0}, y={FER}, col sep=comma] {figs/data/1536_384/1536,384_apc_d.csv};
                
                \coordinate (zoom1) at (axis cs:2.25,6e-6);

        \nextgroupplot[grid=both, grid style={gray!30}, tick align=inside, tickpos=left, yticklabels={}, xticklabel style = {font=\footnotesize}, ymode=log, ymax=1.3,  ymin=2e-6,
                        legend style={font=\footnotesize, nodes={scale=0.7}, at={(0.5,1.03)}, anchor=south, legend columns=3},
                        legend cell align={left},
                        legend entries={~{Multi-kernel}, ~{Short WL}, ~{Punct QUP}, ~{Asymm ASC}, ~{Asymm DES}}
                        ]
                \addplot[Paired-1, mark=square, mark size=1.6pt, thin] table [x={EbN0}, y={FER}, col sep=comma] {figs/data/1536_768/1536,768_mk_h.csv};
                \addplot[Paired-3, mark=*, mark size=1.6pt, thin] table [x={EbN0}, y={FER}, col sep=comma] {figs/data/1536_768/1536,768_s_w.csv};
                \addplot[Paired-5, mark=o, mark size=1.6pt, thin] table [x={EbN0}, y={FER}, col sep=comma] {figs/data/1536_768/1536,768_p_q.csv};
                \addplot[black, mark=triangle, mark size=1.6pt, thin] table [x={EbN0}, y={FER}, col sep=comma] {figs/data/1536_768/1536,768_apc_a.csv};
                \addplot[black, mark=triangle*, mark options={rotate=180}, mark size=1.6pt, thin] table [x={EbN0}, y={FER}, col sep=comma] {figs/data/1536_768/1536,768_apc_d.csv};

                \coordinate (zoom2) at (axis cs:2.5,2e-5);

        \nextgroupplot[grid=both, grid style={gray!30}, tick align=inside, tickpos=left, yticklabels={}, ylabel style = {font=\footnotesize}, xticklabel style = {font=\footnotesize}, ymode=log, ymax=1.3,  ymin=2e-6]
                \addplot[Paired-1, mark=square, mark size=1.6pt, thin] table [x={EbN0}, y={FER}, col sep=comma] {figs/data/1536_1152/1536,1152_mk_h.csv};
                \addplot[Paired-3, mark=*, mark size=1.6pt, thin] table [x={EbN0}, y={FER}, col sep=comma] {figs/data/1536_1152/1536,1152_s_w.csv};
                \addplot[Paired-5, mark=o, mark size=1.6pt, thin] table [x={EbN0}, y={FER}, col sep=comma] {figs/data/1536_1152/1536,1152_p_q.csv};
                \addplot[black, mark=triangle, mark size=1.6pt, thin] table [x={EbN0}, y={FER}, col sep=comma] {figs/data/1536_1152/1536,1152_apc_a.csv};
                \addplot[black, mark=triangle*, mark options={rotate=180}, mark size=1.6pt, thin] table [x={EbN0}, y={FER}, col sep=comma] {figs/data/1536_1152/1536,1152_apc_d.csv};

                \coordinate (zoom3) at (axis cs:3.5,5e-5);


        \nextgroupplot[ylabel=FER, grid=both, grid style={gray!30}, tick align=inside, tickpos=left, yticklabel style = {font=\scriptsize}, ylabel style = {font=\footnotesize},  xticklabel style = {font=\footnotesize}, ymode=log, ymax=1.3,  ymin=2e-6]
                \addplot[Paired-1, mark=square, mark size=1.6pt, thin] table [x={EbN0}, y={FER}, col sep=comma] {figs/data/2304_576/2304,576_mk_h.csv};
                \addplot[Paired-3, mark=*, mark size=1.6pt, thin] table [x={EbN0}, y={FER}, col sep=comma] {figs/data/2304_576/2304,576_s_w.csv};
                \addplot[Paired-5, mark=o, mark size=1.6pt, thin] table [x={EbN0}, y={FER}, col sep=comma] {figs/data/2304_576/2304,576_p_q.csv};
                \addplot[black, mark=triangle, mark size=1.6pt, thin] table [x={EbN0}, y={FER}, col sep=comma] {figs/data/2304_576/2304,576_apc_a.csv};
                \addplot[black, mark=triangle*, mark options={rotate=180}, mark size=1.6pt, thin] table [x={EbN0}, y={FER}, col sep=comma] {figs/data/2304_576/2304,576_apc_d.csv};

                \coordinate (zoom4) at (axis cs:2,6e-6);

        \nextgroupplot[grid=both, grid style={gray!30}, tick align=inside, tickpos=left, yticklabels={}, ylabel style = {font=\footnotesize}, xticklabel style = {font=\footnotesize}, ymode=log, ymax=1.3,  ymin=2e-6]
                \addplot[Paired-1, mark=square, mark size=1.6pt, thin] table [x={EbN0}, y={FER}, col sep=comma] {figs/data/2304_1152/2304,1152_mk_h.csv};
                \addplot[Paired-3, mark=*, mark size=1.6pt, thin] table [x={EbN0}, y={FER}, col sep=comma] {figs/data/2304_1152/2304,1152_s_w.csv};
                \addplot[Paired-5, mark=o, mark size=1.6pt, thin] table [x={EbN0}, y={FER}, col sep=comma] {figs/data/2304_1152/2304,1152_p_q.csv};
                \addplot[black, mark=triangle, mark size=1.6pt, thin] table [x={EbN0}, y={FER}, col sep=comma] {figs/data/2304_1152/2304,1152_apc_a.csv};
                \addplot[black, mark=triangle*, mark options={rotate=180}, mark size=1.6pt, thin] table [x={EbN0}, y={FER}, col sep=comma] {figs/data/2304_1152/2304,1152_apc_d.csv};

                \coordinate (zoom5) at (axis cs:2.25,4e-5);

        \nextgroupplot[grid=both, grid style={gray!30}, tick align=inside, tickpos=left, yticklabels={}, ylabel style = {font=\footnotesize}, xticklabel style = {font=\footnotesize}, ymode=log, ymax=1.3,  ymin=2e-6]
                \addplot[Paired-1, mark=square, mark size=1.6pt, thin] table [x={EbN0}, y={FER}, col sep=comma] {figs/data/2304_1728/2304,1728_mk_h.csv};
                \addplot[Paired-3, mark=*, mark size=1.6pt, thin] table [x={EbN0}, y={FER}, col sep=comma] {figs/data/2304_1728/2304,1728_s_w.csv};
                \addplot[Paired-5, mark=o, mark size=1.6pt, thin] table [x={EbN0}, y={FER}, col sep=comma] {figs/data/2304_1728/2304,1728_p_q.csv};
                \addplot[black, mark=triangle, mark size=1.6pt, thin] table [x={EbN0}, y={FER}, col sep=comma] {figs/data/2304_1728/2304,1728_apc_a.csv};
                \addplot[black, mark=triangle*, mark options={rotate=180}, mark size=1.6pt, thin] table [x={EbN0}, y={FER}, col sep=comma] {figs/data/2304_1728/2304,1728_apc_d.csv};

                \coordinate (zoom6) at (axis cs:3.25,1.7e-4);


        \nextgroupplot[xlabel= \footnotesize {${E_b/N_0}$ (dB)}, ylabel=FER, grid=both, grid style={gray!30}, tick align=inside, tickpos=left, yticklabel style = {font=\scriptsize}, ylabel style = {font=\footnotesize},  xticklabel style = {font=\footnotesize}, ymode=log, ymax=1.3,  ymin=2e-6]
                \addplot[Paired-1, mark=square, mark size=2pt, thin] table [x={EbN0}, y={FER}, col sep=comma] {figs/data/3072_768/3072,768_mk_h.csv};
                \addplot[Paired-3, mark=*, mark size=2pt, thin] table [x={EbN0}, y={FER}, col sep=comma] {figs/data/3072_768/3072,768_s_w.csv};
                \addplot[Paired-5, mark=o, mark size=2pt, thin] table [x={EbN0}, y={FER}, col sep=comma] {figs/data/3072_768/3072,768_p_q.csv};
                \addplot[black, mark=triangle, mark size=2pt, thin] table [x={EbN0}, y={FER}, col sep=comma] {figs/data/3072_768/3072,768_apc_a.csv};
                \addplot[black, mark=triangle*, mark options={rotate=180}, mark size=2pt, thin] table [x={EbN0}, y={FER}, col sep=comma] {figs/data/3072_768/3072,768_apc_d.csv};

                \coordinate (zoom7) at (axis cs:1.75,1.6e-5);

        \nextgroupplot[xlabel= \footnotesize {${E_b/N_0}$ (dB)}, grid=both, grid style={gray!30}, tick align=inside, tickpos=left, yticklabels={}, ylabel style = {font=\footnotesize}, xticklabel style = {font=\footnotesize}, ymode=log, ymax=1.3,  ymin=2e-6]
                \addplot[Paired-1, mark=square, mark size=2pt, thin] table [x={EbN0}, y={FER}, col sep=comma] {figs/data/3072_1536/3072,1536_mk_h.csv};
                \addplot[Paired-3, mark=*, mark size=2pt, thin] table [x={EbN0}, y={FER}, col sep=comma] {figs/data/3072_1536/3072,1536_s_w.csv};
                \addplot[Paired-5, mark=o, mark size=2pt, thin] table [x={EbN0}, y={FER}, col sep=comma] {figs/data/3072_1536/3072,1536_p_q.csv};
                \addplot[black, mark=triangle, mark size=2pt, thin] table [x={EbN0}, y={FER}, col sep=comma] {figs/data/3072_1536/3072,1536_apc_a.csv};
                \addplot[black, mark=triangle*, mark options={rotate=180}, mark size=2pt, thin] table [x={EbN0}, y={FER}, col sep=comma] {figs/data/3072_1536/3072,1536_apc_d.csv};

                \coordinate (zoom8) at (axis cs:2.25,1e-5);

        \nextgroupplot[xlabel= \footnotesize {${E_b/N_0}$ (dB)}, grid=both, grid style={gray!30}, tick align=inside, tickpos=left, yticklabels={}, ylabel style = {font=\footnotesize}, xticklabel style = {font=\footnotesize}, ymode=log, ymax=1.3,  ymin=2e-6]
                \addplot[Paired-1, mark=square, mark size=2pt, thin] table [x={EbN0}, y={FER}, col sep=comma] {figs/data/3072_2304/3072,2304_mk_h.csv};
                \addplot[Paired-3, mark=*, mark size=2pt, thin] table [x={EbN0}, y={FER}, col sep=comma] {figs/data/3072_2304/3072,2304_s_w.csv};
                \addplot[Paired-5, mark=o, mark size=2pt, thin] table [x={EbN0}, y={FER}, col sep=comma] {figs/data/3072_2304/3072,2304_p_q.csv};
                \addplot[black, mark=triangle, mark size=2pt, thin] table [x={EbN0}, y={FER}, col sep=comma] {figs/data/3072_2304/3072,2304_apc_a.csv};
                \addplot[black, mark=triangle*, mark options={rotate=180}, mark size=2pt, thin] table [x={EbN0}, y={FER}, col sep=comma] {figs/data/3072_2304/3072,2304_apc_d.csv};

                \coordinate (zoom9) at (axis cs:3.25,7e-5);
\end{groupplot}

\spy [black, height=1.9cm, width=0.7cm] on (zoom1) in node [fill=white] at (0.8,1.4);
\spy [black, height=1.1cm, width=0.7cm] on (zoom2) in node [fill=white] at (3.2,1);
\spy [black, height=1.9cm, width=0.7cm] on (zoom3) in node [fill=white] at (5.6,1.4);

\spy [black, height=2.1cm, width=0.7cm] on (zoom4) in node [fill=white] at (0.8,-3.6);
\spy [black, height=1.2cm, width=0.7cm] on (zoom5) in node [fill=white] at (3.2,-4);
\spy [black, height=1.9cm, width=0.7cm] on (zoom6) in node [fill=white] at (5.6,-3.7);

\spy [black, height=1.8cm, width=0.7cm] on (zoom7) in node [fill=white] at (0.8,-8.9);
\spy [black, height=1.1cm, width=0.7cm] on (zoom8) in node [fill=white] at (3.2,-9.25);
\spy [black, height=2.1cm, width=0.7cm] on (zoom9) in node [fill=white] at (5.6,-8.7);

\node[fill = white, draw=black, inner sep=0.5mm, above left = 0.1cm and -0.8cm of fer_group c1r1.south] {\tiny (a) $\mathcal{PC}(1536,384)$};
\node[fill = white, draw=black, inner sep=0.5mm, above left = 0.1cm and -0.8cm of fer_group c2r1.south] {\tiny (b) $\mathcal{PC}(1536,768)$};
\node[fill = white, draw=black, inner sep=0.5mm, above left = 0.1cm and -0.8cm of fer_group c3r1.south] {\tiny (c) $\mathcal{PC}(1536,1152)$};

\node[fill = white, draw=black, inner sep=0.5mm, above left = 0.1cm and -0.8cm of fer_group c1r2.south] {\tiny (d) $\mathcal{PC}(2304,576)$};
\node[fill = white, draw=black, inner sep=0.5mm, above left = 0.1cm and -0.8cm of fer_group c2r2.south] {\tiny (e) $\mathcal{PC}(2304,1152)$};
\node[fill = white, draw=black, inner sep=0.5mm, above left = 0.1cm and -0.8cm of fer_group c3r2.south] {\tiny (f) $\mathcal{PC}(2304,1728)$};

\node[fill = white, draw=black, inner sep=0.5mm, above left = 0.1cm and -0.8cm of fer_group c1r3.south] {\tiny (g) $\mathcal{PC}(3072,768)$};
\node[fill = white, draw=black, inner sep=0.5mm, above left = 0.1cm and -0.8cm of fer_group c2r3.south] {\tiny (h) $\mathcal{PC}(3072,1536)$};
\node[fill = white, draw=black, inner sep=0.5mm, above left = 0.1cm and -0.8cm of fer_group c3r3.south] {\tiny (i) $\mathcal{PC}(3072,2304)$};
\end{tikzpicture}

%% file: figs/fssc_ops_group_plot.tikz
\begin{tikzpicture}

    \begin{groupplot}[
        group style={group name=fssc_group, group size= 2 by 1, horizontal sep=7mm, vertical sep=10mm,}, height=0.58\linewidth, width=0.58\linewidth]

        \nextgroupplot[
        ylabel=Number of operations,
        ymode=log, log basis y = 2,
        xtick={0.125,0.25,0.375,0.5,0.625,0.75,0.875},
        xticklabels={$\frac{1}{8}$,$\frac{2}{8}$,$\frac{3}{8}$,$\frac{4}{8}$,$\frac{5}{8}$,$\frac{6}{8}$,$\frac{7}{8}$},
        yticklabel style={rotate=0, font=\scriptsize},
        ymajorgrids]
                \addplot[Paired-3, mark=*, mark size=1.6pt, only marks] table [x={R}, y={S1}, col sep=space] {figs/data/K164_WL_64.dat}; \label{plots:wl}
                \addplot[Paired-5, mark=o, mark size=1.6pt, only marks] table [x={R}, y={S1}, col sep=space] {figs/data/K164_QUP_64.dat}; \label{plots:qup}
                \addplot[black, mark=triangle, mark size=1.6pt, only marks] table [x={R}, y={S1}, col sep=space] {figs/data/K164_ASC_64.dat}; \label{plots:asc}
                \addplot[black, mark=triangle*, mark options={rotate=180}, mark size=1.6pt, only marks] table [x={R}, y={S1}, col sep=space] {figs/data/K164_DEC_64.dat}; \label{plots:dec}
                \coordinate (top) at (rel axis cs:0,1);

        \nextgroupplot[
        xtick={0.125,0.25,0.375,0.5,0.625,0.75,0.875},
        xticklabels={$\frac{1}{8}$,$\frac{2}{8}$,$\frac{3}{8}$,$\frac{4}{8}$,$\frac{5}{8}$,$\frac{6}{8}$,$\frac{7}{8}$},
        yticklabel style={rotate=0, font=\scriptsize},
        ymajorgrids]
                \addplot[Paired-3, mark=*, mark size=1.6pt, only marks] table [x={R}, y={S1}, col sep=space] {figs/data/N768_WL_64.dat};
                \addplot[Paired-5, mark=o, mark size=1.6pt, only marks] table [x={R}, y={S1}, col sep=space] {figs/data/N768_QUP_64.dat};
                \addplot[black, mark=triangle, mark size=1.6pt, only marks] table [x={R}, y={S1}, col sep=space] {figs/data/N768_ASC_64.dat};
                \addplot[black, mark=triangle*, mark options={rotate=180}, mark size=1.6pt, only marks] table [x={R}, y={S1}, col sep=space] {figs/data/N768_DEC_64.dat};

                \coordinate (bot) at (rel axis cs:1,0);

    \end{groupplot}

    \path (fssc_group c1r1.south east) --node[below=0.4cm]{$R$} (fssc_group c2r1.south west);

    \path (top|-current bounding box.north) -- 
      coordinate(legendpos)
      (bot|-current bounding box.north);
    \matrix[
        matrix of nodes,
        anchor=south,
        draw,
        inner sep=0.2em,
        draw
      ]at([yshift=1ex]legendpos)
      {
        \ref{plots:wl}& \footnotesize Short WL &[5pt]
        \ref{plots:qup}& \footnotesize Punct QUP &[5pt]
        \ref{plots:asc}& \footnotesize Asymm ASC &[5pt]
        \ref{plots:dec}& \footnotesize Asymm DES  \\};

    \node[fill = white, draw=black, inner sep=0.5mm, above left = 0.15cm and -0.8cm of fssc_group c1r1.south] {\footnotesize (a) $K=164$};
    \node[fill = white, draw=black, inner sep=0.5mm, above left = 0.15cm and -0.8cm of fssc_group c2r1.south] {\footnotesize (b) $N=768$};

\end{tikzpicture}

%% file: bare_conf.bbl
\begin{thebibliography}{10}
\providecommand{\url}[1]{#1}
\csname url@samestyle\endcsname
\providecommand{\newblock}{\relax}
\providecommand{\bibinfo}[2]{#2}
\providecommand{\BIBentrySTDinterwordspacing}{\spaceskip=0pt\relax}
\providecommand{\BIBentryALTinterwordstretchfactor}{4}
\providecommand{\BIBentryALTinterwordspacing}{\spaceskip=\fontdimen2\font plus
\BIBentryALTinterwordstretchfactor\fontdimen3\font minus
  \fontdimen4\font\relax}
\providecommand{\BIBforeignlanguage}[2]{{%
\expandafter\ifx\csname l@#1\endcsname\relax
\typeout{** WARNING: IEEEtran.bst: No hyphenation pattern has been}%
\typeout{** loaded for the language `#1'. Using the pattern for}%
\typeout{** the default language instead.}%
\else
\language=\csname l@#1\endcsname
\fi
#2}}
\providecommand{\BIBdecl}{\relax}
\BIBdecl

\bibitem{Arikan2009}
E.~Arikan, ``Channel {P}olarization: {A} {M}ethod for {C}onstructing
  {C}apacity-{A}chieving {C}odes for {S}ymmetric {B}inary-{I}nput {M}emoryless
  {C}hannels,'' \emph{{IEEE} Trans. on Inform. Theory}, vol.~55, no.~7, pp.
  3051--3073, Jul. 2009.

\bibitem{3GPP}
{3GPP}, ``{NR; Multiplexing and Channel Coding},''
  \url{{http://www.3gpp.org/DynaReport/38-series.htm}}, Tech. Rep. TS 38.212,
  June 2018, {Release 15}.

\bibitem{Richardson2018}
T.~Richardson and S.~Kudekar, ``Design of {L}ow-{D}ensity {P}arity {C}heck
  {C}odes for {5G} {N}ew {R}adio,'' \emph{{IEEE} Communications Magazine},
  vol.~56, no.~3, pp. 28--34, Mar. 2018.

\bibitem{Niu2013}
K.~Niu, K.~Chen, and J.-R. Lin, ``Beyond {T}urbo {C}odes: {R}ate-{C}ompatible
  {P}unctured {P}olar {C}odes,'' in \emph{2013 {IEEE} Int. Conf. Commun.
  ({ICC})}, 2013.

\bibitem{Wang2014}
R.~Wang and R.~Liu, ``A {Novel} {P}uncturing {S}cheme for {P}olar {C}odes,''
  \emph{{IEEE} Commun. Lett.}, vol.~18, no.~12, pp. 2081--2084, Dec. 2014.

\bibitem{Gabry2016}
F.~Gabry, V.~Bioglio \emph{et~al.}, ``Multi-{K}ernel {C}onstruction of {P}olar
  {C}odes,'' \emph{{IEEE} Int. Conf. Commun. ({ICC})}, 2017.

\bibitem{7938042}
P.~Trifonov, ``Chained {P}olar {S}ubcodes,'' in \emph{SCC 2017; 11th Int. ITG
  Conf. on Syst., Commun. and Coding}, Feb. 2017, pp. 1--6.

\bibitem{Bioglio2017}
V.~Bioglio, F.~Gabry, and I.~Land, ``Low-{C}omplexity {P}uncturing and
  {S}hortening of {P}olar {C}odes,'' in \emph{2017 {IEEE} Wireless Commun. and
  Networking Conf. Workshops ({WCNCW})}.\hskip 1em plus 0.5em minus 0.4em\relax
  {IEEE}, Mar. 2017.

\bibitem{Benammar2017}
M.~Benammar, V.~Bioglio \emph{et~al.}, ``Multi-{K}ernel {P}olar {C}odes:
  {P}roof of {P}olarization and {E}rror {E}xponents,'' \emph{{IEEE} Info.
  Theory Workshop}, 2017.

\bibitem{Coppolino2018}
G.~Coppolino, C.~Condo \emph{et~al.}, ``A {M}ulti-{K}ernel {M}ulti-{C}ode
  {P}olar {D}ecoder {A}rchitecture,'' \emph{{IEEE} Trans. on Circuits and Syst.
  I: Regular Papers}, pp. 1--10, 2018.

\bibitem{Tal2015}
I.~Tal and A.~Vardy, ``List {D}ecoding of {P}olar {C}odes,'' \emph{{IEEE}
  Trans. on Inform. Theory}, vol.~61, no.~5, pp. 2213--2226, May 2015.

\bibitem{Sarkis2014}
G.~Sarkis, P.~Giard \emph{et~al.}, ``Fast {P}olar {D}ecoders: {A}lgorithm and
  {I}mplementation,'' \emph{{IEEE} J. on Select. Areas in Commun.}, vol.~32,
  no.~5, pp. 946--957, May 2014.

\bibitem{Sarkis2016}
G.~Sarkis, P.~Giard \emph{et~al.}, ``Fast {L}ist {D}ecoders for {P}olar
  {C}odes,'' \emph{{IEEE} J. on Select. Areas in Commun.}, vol.~34, no.~2, pp.
  318--328, Feb. 2016.

\bibitem{Trifonov2012}
P.~Trifonov, ``Efficient {D}esign and {D}ecoding of {P}olar {C}odes,''
  \emph{{IEEE} Trans. on Commun.}, vol.~60, no.~11, pp. 3221--3227, Nov. 2012.

\bibitem{Cassagne2017a}
A.~Cassagne, O.~Hartmann \emph{et~al.}, ``Fast {S}imulation and {P}rototyping
  with {AFF3CT},'' in \emph{Int. Workshop on Signal Proces. Syst.}\hskip 1em
  plus 0.5em minus 0.4em\relax {IEEE}, Oct. 2017.

\bibitem{Wu2014}
D.~Wu, Y.~Li, and Y.~Sun, ``Construction and {B}lock {E}rror {R}ate {A}nalysis
  of {P}olar {C}odes {O}ver {AWGN} {C}hannel {B}ased on {G}aussian
  {A}pproximation,'' \emph{{IEEE} Commun. Letters}, vol.~18, no.~7, pp.
  1099--1102, Jul. 2014.

\bibitem{Leroux2013}
C.~Leroux, A.~J. Raymond \emph{et~al.}, ``A {S}emi-{P}arallel
  {S}uccessive-{C}ancellation {D}ecoder for {P}olar {C}odes,'' \emph{{IEEE}
  Trans. on Signal Processing}, vol.~61, no.~2, pp. 289--299, Jan. 2013.

\end{thebibliography}
